\documentclass[pre,amsmath,amssymb,twocolumn,showpacs,superscriptaddress,longbibliography]{revtex4-1}
\usepackage{amsfonts,amsmath,amssymb,bm}
\usepackage{graphicx,graphics,float}
\usepackage{bm}
\usepackage{amssymb}
\usepackage{colordvi}
\usepackage{graphicx}
\usepackage{color}
\usepackage[colorlinks=true,linkcolor=blue,citecolor=blue,urlcolor=blue]{hyperref}
\usepackage{hyperref}
\usepackage{comment}
\usepackage{harpoon}
\usepackage{braket}
\usepackage{booktabs}
\usepackage{array}

\begin{document}

\title{Multitask quantum thermal machines and cooperative effects}

\author{Jincheng Lu}
\address{Jiangsu Key Laboratory of Micro and Nano Heat Fluid Flow Technology and Energy Application, School of Physical Science and Technology, Suzhou University of Science and Technology, Suzhou, 215009, China}


\author{Zi Wang}
\address{Center for Phononics and Thermal Energy Science, China-EU Joint Lab on Nanophononics, Shanghai Key Laboratory of Special Artificial Microstructure Materials and Technology, School of Physics Science and Engineering, Tongji University, Shanghai 200092, China}

\author{Rongqian Wang}
\affiliation{Institute of theoretical and applied physics \& School of physical science and technology \&
Collaborative Innovation Center of Suzhou Nano Science and Technology, Soochow University, Suzhou 215006, China.}

\author{Jiebin Peng}
\address{School of Physics and Optoelectronic Engineering, Guangdong University of Technology, Guangzhou 510006, Guangdong Province, China}


\author{Chen Wang}\email{Corresponding author: wangchen@zjnu.cn}
\address{Department of Physics, Zhejiang Normal University, Jinhua, Zhejiang 321004, China}

\author{Jian-Hua Jiang}\email{Corresponding author: jianhuajiang@suda.edu.cn}
\affiliation{Institute of theoretical and applied physics \& School of physical science and technology \&
Collaborative Innovation Center of Suzhou Nano Science and Technology, Soochow University, Suzhou 215006, China.}

\date{\today}


\begin{abstract}
Including phonon-assisted inelastic process in thermoelectric devices is able to enhance the performance of nonequilibrium work extraction.
In this work, we demonstrate that inelastic phonon-thermoelectric devices have a fertile functionality diagram,
where particle current and phononic heat currents are coupled and fueled by chemical potential difference.
Such devices can simultaneously perform multiple tasks, e.g., heat engines, refrigerators, and heat pumps.
Guided by the entropy production, we mainly study the efficiencies and coefficients of performance of multitask quantum thermal machines,
where the roles of the inelastic scattering process and multiple biases in multiterminal setups are
emphasized.
Specifically, in a three-terminal double-quantum-dot setup with a tunable gate, we show that it efficiently performs two useful tasks due to the phonon-assisted inelastic process.
Moreover, the cooperation between the longitudinal and transverse thermoelectric effects in the three-terminal thermoelectric systems leads to markedly improved performance of the thermal machines.
While for the four-terminal four-quantum-dot thermoelectric setup,
we find that additional thermodynamic affinity furnishes the system with both  enriched functionality  and enhanced efficiency.
Our work provides insights into optimizing phonon-thermoelectric devices.
\end{abstract}


\maketitle

\section{Introduction}
Understanding the performance of thermoelectric transport at the nanoscale is of significant importance for its potential applications in quantum technology~\cite{Nanotechnology,JiangCRP,BenentiPR17,RMPPekola,MyReview}.
 Much effort has been devoted to achieving high efficiency and coefficient of performance in thermoelectric nanodevices both theoretically~\cite{Mukherjee21,WhitneyPRL,WhitneyPRB,JiangOra} and experimentally\cite{Josefsson,zhai22},
which mainly originate from the elastic transport processes~\cite{Proesmans,Naoto,YuPRE19,Constancy,MaPRL20,YuEntropy}.
Interestingly, Mahan and Sofo proposed the ``best thermoelectrics'' by using conductors with very narrow energy bands to reduce the thermal conductivity based on the Wiedemann-Franz law~\cite{Mahan}.
However, it was recently discovered that phonon thermal transport will inevitably suppress the figure or merit, and output power in the zero band width limit~\cite{ZhouPRL}. The carrier of particle and heat current for the elastic transport processes is electrons, so it is impossible to spatially separate them.
In contrast, it may be available to realize such separation of heat and charge currents in inelastic transport processes~\cite{MazzaPRB}.

Recently, another fundamental category, i.e., inelastic transport processes, has been explored~\cite{OraPRB2010,DavidPRL,JiangNJP,SothmannQW,Yamamoto,ChenPRE15,BijayJiang,JiangBijayPRB17,WangPRL}.
One exciting perspective is that the inclusion of an additional terminal to ``decouple'' the particle and heat currents may significantly improve thermoelectric efficiency.
Until now, a vast of valuable works have been carried out to study inelastic transport in thermoelectric systems, e.g., Coulomb-coupled quantum dot (QD) systems~\cite{Rafael,ThiersNJP2015}, metal-superconductor junctions~\cite{TabatabaeiPRL20}, and boson-assisted thermoelectric devices~\cite{Jiang2012,Jiang2013,Cooling2,MyPRBdiode}.
However, the study of multiterminal energy conversion and transfer is still in its infancy, in which the thermodynamic machine usually performs a single task, such as refrigerators~\cite{ReviewKosloff,Rongqian,David-refrigerator}, heat pumps~\cite{RenPRL10,wangpump,WangZiPRE,AVS}, or the heat engines~\cite{ManzanoReview,tu21,Jiang2017,JiangNearfield,MyPRBGTUR}.
Intriguingly, as proposed by Entin-Wohlman {\it et al.}~\cite{Ora2015} and  Manzano {\it et al.}~\cite{ManzanoPRR}, a three-terminal device is considered as a hybrid thermal machine,
which performs multiple efficient tasks simultaneously by introducing a reference temperature.
Consequently, the thermoelectric effect accompanied by one electric current and two heat currents, is analyzed.

In this work,  we give a universal exergy efficiency without relying on the particular reference temperature,
based on the concept of entropy production.
We show that inelastic thermoelectric devices with multiple electric currents and/or multiple heat currents can be nominated as hybrid thermal machines~\cite{hammam22}, i.e., devices concurrently performing multiple useful tasks.
Specifically, we desire to analyze different thermoelectric devices performing distinct tasks,
 e.g., heat engines, refrigerators, and heat pumps.
When a system connects to multiple reservoirs,
the entropy production rate of the whole system is described as ${\dot S}_{\rm tot} = \sum_{i=1}^N{\dot S}_i$, where ${\dot S}_i$ is the entropy production rate of the $i$th reservoir~\cite{RMPLandi}.
If we denote $T$ as the temperature of a reservoir,
 which is the intermediate temperature compared with others (instead of deliberately introducing a reference temperature),
the entropy production can be decomposed into~\cite{Thomas92,RMPLandi}
\begin{equation}
T{\dot S}_{\rm tot}= T\sum_{v=1}^N{\dot S}_v = \sum_{v=1}^N I_vA_v,
\label{eq:Stot}
\end{equation}
where $I_v$ is the current and $A_v$ is the corresponding affinity.
Then, the exergy efficiency is defined as the ratio of all the entropy decrease terms (the inputs) to
 all the entropy increase terms (the outputs)~\cite{ManzanoPRR,MukherjeeJAP,HajilooPRB,CarregaPRXQ}
\begin{equation}
\phi = -\frac{\sum_v^- {\dot S}_v }{\sum_v^+ {\dot S}_v}\le 1,
\label{eq:phi}
\end{equation}
where $\sum_v^{\pm}S_v$ denotes the sum over the positive and negative terms of the rates, respectively.
According to the second law of thermodynamics, the exergy efficiency is bounded via $\phi\le 1$~\cite{PRXQ}.
In Eqs.~\eqref{eq:Stot} and \eqref{eq:phi}, we use the temperature of a reservoir $T$ in the middle
to characterize whether a process is negative entropy production (useful) or positive entropy production (useless). In this way, we avoid the situations that different choices of reference temperature lead to different conclusions, and may even violate the second law of thermodynamics.

\begin{figure}[htb]
\begin{center}
\centering\includegraphics[width=8.5cm]{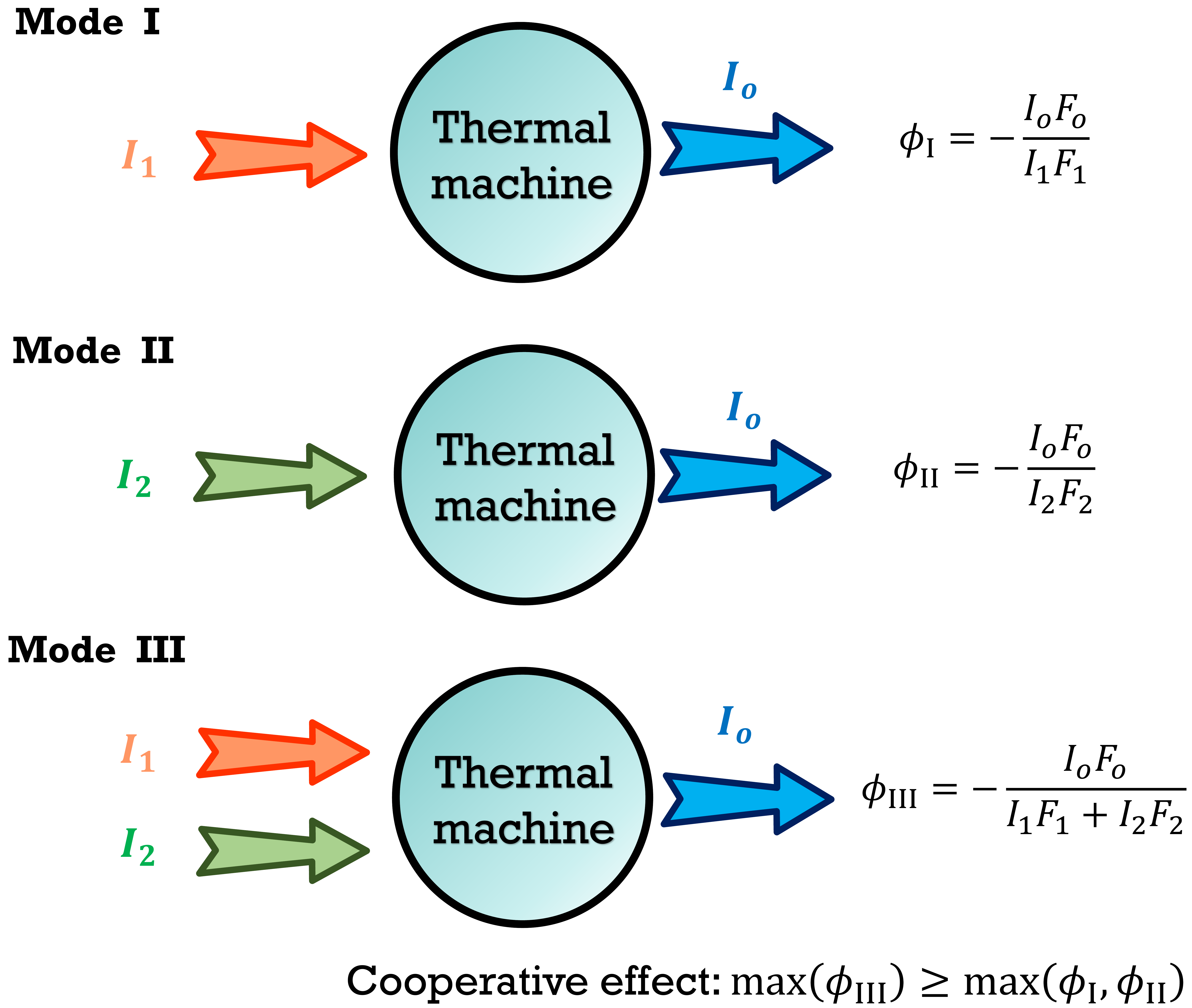}
\caption{ Schematic illustration of thermoelectric cooperative effects in quantum thermal machine. $I_i$ ($i=1,2$) denotes the input current and $I_o$ is the output current, $F_i$ is the corresponding force. $\phi_{i}$ ($i={\rm I,II,II}$) is the exergy efficiency of the thermal machine. }
\label{fig:coopera}
\end{center}
\end{figure}


The exergy efficiency [Eq.~\eqref{eq:phi}] has been extensively studied both at the equilibrium and nonequilibrium regimes~\cite{JiangPRE,Snyder,HajilooPRB}. Interestingly, the relation between the exergy efficiency $\phi$ and conventional efficiency $\eta=W/Q$ ($W$ and $Q$ are the output work and input heat) is obtained as~\cite{JiangPRE}
\begin{eqnarray}
\phi=\eta/\eta_C,
\end{eqnarray}
where $\eta_C=1-T_c/T_h$ is the Carnot efficiency, and $T_c$ and $T_h$ are the temperatures of the cold and hot reservoirs, respectively.
The important issue in quantum thermodynamics is to find out the optimal efficiency and the corresponding maximal power of thermal machines.

While considering energy conversion,
thermoelectric cooperative effects have been demonstrated to be an effectual way to improve conversion efficiency (figure of merit)~\cite{JiangPRE,JiangJAP,MyJAP,CPB}. As shown in Fig.~\ref{fig:coopera}, the thermal machine connects to two input currents and one output current.
 And there are two thermoelectric effects associated with different chemical potentials and/or temperature gradients. The essence of cooperative effect is to modulate the input currents through regulating the thermodynamic forces. For instance, there are only one input current in mode I and mode II, while mode III has two input currents. It is found that the exergy efficiency of mode III can be greater than mode I and II, i.e., $\max(\phi_{\rm III})\ge \max{(\phi_{\rm I},\phi_{\rm II})}$.

In the work, we primarily examine the general theoretical framework of multiterminal devices,
i.e., double QDs three-terminal system
comprising two electronic terminals and a phonon bath (Sec.~\ref{secDQD}) and four-QD four-terminal system (Sec.~\ref{sec4QD}), defines the particle and heat currents and the corresponding thermodynamic driving forces, and uses those quantities to reexpress the efficiencies of the thermal machine for different tasks.
We also discuss thermodynamic multitasks  of the various setups in the linear response, based on the Onsager formalism.
Finally section \ref{conclusion} is devoted to the conclusions.
We also discuss the performance of three-terminal single-level QD system that perform multiple useful tasks simultaneously in Appendix~\ref{sec1QD}.
Throughout this work, we set electron charge $e\equiv1$ and Planck constant $\hbar\equiv1$.

\section{The three-terminal double QDs device}\label{secDQD}
In this section we will present a generic inelastic transport model to  illustrate multitasks of hybrid thermal machines.
The simplest nontrivial geometry configuration for inelastic transport is the three-terminal system, where energy absorbed/emitted by the electron is assisted by a third phonon bath, differing from the left reservoir and right reservoir.
This model is characterized as a hopping double QDs model~\cite{Jiang2012,Jiangtransistors,Jiang2017,MyPRBdiode,MyPRBtransistor}.
At linear thermoelectric response regime, one remarkable feature of double QDs is that the three-terminal system exhibits a large thermopower and high figure of merit~\cite{Jiang2012}.
Moreover, it is found that nonlinear thermoelectric transport can further enhance energy efficiency and output power under nonlinear driving sources~\cite{Jiang2017}.
Here, we mainly focus on the efficiencies and coefficients of performance of multi-task three-terminal  inelastic phonon-thermoelectric device.

\subsection{The basic setup and currents}

\begin{figure}[htb]
\begin{center}
\centering\includegraphics[width=8.5cm]{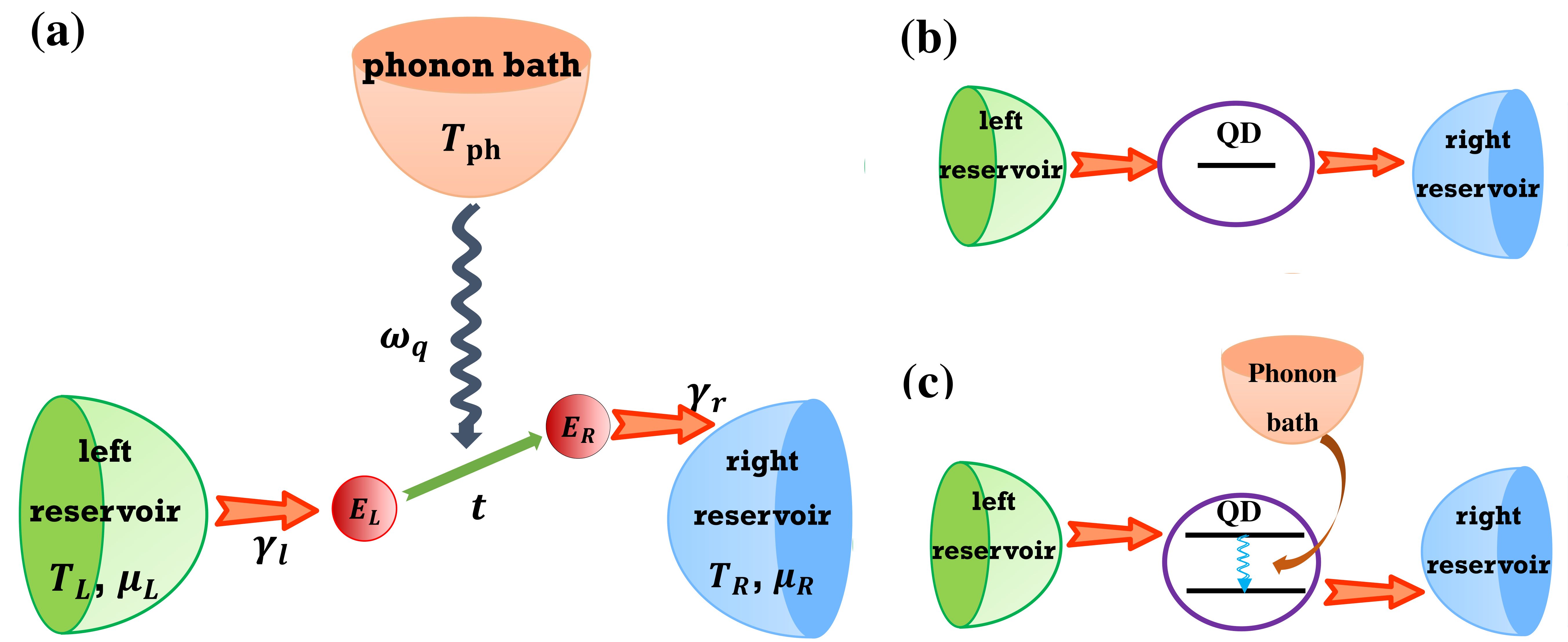}
\caption{Illustration of three-terminal double QD system. An electron left the left reservoir into the left QD (with energy $E_L$) hops to the right QD (with a energy $E_R$) as assisted by a phonon from the phonon bath (with temperature $T_{\rm ph}$). The electron then tunnels into the right reservoir from the right QD. The electrochemical potential and temperature of the left reservoir (right reservoir) are $\mu_L$ and $T_L$ ($\mu_R$ and $T_R$), respectively. $\gamma_{l/r}$ are the hybridization energies of the QDs to the left/right reservoir, respectively. Illustration of (b) elastic, (c) inelastic thermoelectric transport. }~\label{fig:DQD}
\end{center}
\end{figure}

The three-terminal double QDs device consists of the left reservoir, the right reservoir, and a phonon bath,
as schematically depicted in Fig.~\ref{fig:DQD}.
The QDs are labeled by $L$ and $R$ with tunable energy levels $E_L$ and $E_R$. An electron leaves the left QD and hops to the right QD, which is simultaneously assisted by one phonon from the phonon bath (with temperature $T_{\rm ph}$).  The Hamiltonian of the double-QDs device is $\hat H = \hat H_{\rm DQD} + \hat H_{\rm e-ph} + \hat H_{\rm lead} + \hat H_{\rm tun} + \hat H_{\rm ph}$, with $\hat H_{\rm DQD} = \sum_{i=L,R} E_i \hat b_i^\dagger \hat b_i +  t(\hat b_l^\dagger \hat b_r + {\rm H.c.})$, $\hat H_{\rm e-ph} = \sum_{q}\lambda_q \hat b_l^\dagger \hat b_r (\hat d_q + \hat d^\dagger_q) + {\rm H.c.} $, $\hat H_{\rm ph} = \sum_q\omega_{q}\hat d^\dagger_q \hat d_q$, $\hat H_{\rm lead} = \sum_{j=L, R}\sum_{k} \varepsilon_{j,k} \hat b_{j,k}^\dagger \hat b_{j,k}$, and $\hat H_{\rm tun} = \sum_k V_{L, k} \hat b_l^\dagger \hat b_{L, k} + \sum_k V_{R, k} \hat b_r^\dagger \hat b_{R, k} +  {\rm H.c.}$,
where $\hat b_i^\dagger$ ($\hat b_i$) is the creation  (annihilation) operator of electron in the $i$ QD, and  $E_{i}$ is the QD energy, $t$ is the tunneling between the two QDs, $\gamma_{L}$ ($\gamma_{R}$) is the coupling between the QD and the left (right) reservoir, $\lambda_q$ is the strength of electron-phonon interaction, and $\hat d^\dagger_q$~($\hat d_q$) is the creation  (annihilation) operator of phonon with the frequency $\omega_{q}$.

There exist particle $I_p^i$ ($i=L,R$), energy $I_E^i$ ($i=L,R,\rm ph$), and heat $I_Q^i$ ($i=L,R,\rm ph$) currents  out of the  $i$-th reservoir, which are denoted by grey lines with arrows in Fig.~\ref{fig:DQD}~\cite{OraPRB2010,Cooling2}.
Accordingly, the electric and phononic heat currents are given by $I_Q^i = I_E^i- \mu_i I_p^i$ ($i=L,R$) and $I_Q^{\rm ph}=I_E^{\rm ph}$.
We define that the current $I_i$ is positive when flowing towards the hybrid quantum system.
Moreover, the energy conversion leads to the relation $I_E^L + I_E^R + I_Q^{\rm ph} = 0$~\cite{OraPRB2010},
while the particles conversion results in $I_p^L + I_p^R = 0$.
Moreover, The entropy production relation of the  system is given by~\cite{gchen2005book}
$-\dot{S}_{\rm tot} = {I_Q^L}/{T_L} +  {I_Q^R}/{T_R} + {I_Q^{\rm ph}}/{T_{\rm ph}}$.
Combining the particle and energy currents conservation relations, the entropy production rate can be reexpressed as
\begin{equation}
T_L\frac{dS_{\rm tot}}{dt} = I_p^R A_p^R  +  I_Q^R A_Q^R + I_Q^{\rm ph}A_Q^{\rm ph},
\label{eq:dSdt}
\end{equation}
where $A_p^R=\mu_R-\mu_L$, $A_Q^R = 1- \frac{T_L}{T_R}$, and $A_Q^{\rm ph} = 1 - \frac{T_L}{T_{\rm ph}}$.
Based on the Fermi golden rule~\cite{Jiang2012},
one is able to obtain the contribution of inelastic processes to the heat currents
\begin{equation}
\begin{aligned}
&I_Q^{L,\rm inel} = (E_L-\mu_L)I_p^L, \\
&I_Q^{R,\rm inel} = -(E_R-\mu_R)I_p^L, \\
&I_Q^{\rm ph,\rm inel} = (E_R-E_L) I_p^L.
\label{eq:currents}
\end{aligned}
\end{equation}
Here $I_p^{L,\rm inel} = \Gamma_{l\rightarrow r} - \Gamma_{r\rightarrow l}$ with $\Gamma_{l\rightarrow r} \equiv \gamma_{e-{\rm ph}} f_l ( 1 - f_r) N_p^-$ and $\Gamma_{r\rightarrow l} \equiv \gamma_{e-{\rm ph}} f_r (1-f_l) N_p^+$. $N_p^{\pm}=N_B+\frac{1}{2}\pm\frac{1}{2}{\rm sgn}(E_R-E_L)$ with  $N_B\equiv [\exp({|E_R-E_L|}/{T_{\rm ph}})-1]^{-1}$ being the Bose-Einstein distribution for phonons. $\gamma_{e-{\rm ph}}=|\lambda_q|^2v_{\rm ph}$, where $v_{\rm ph}$ denotes the phonon density of states. In the following, we assume that $f_{i}\approx {\exp[(E_{i}-\mu_i)/k_BT_i]+1}^{-1}$, where $f_i$ ($i = L, R$) is the Fermi distribution function for the $i$-th electric reservoir. In Fig.~\ref{fig:DQD}(c), we can find that the inelastic process mainly results from the electron-phonon interaction processes, which involve the collaborative transport between  three  reservoirs.

In contrast, the contribution of elastic processes [see Fig.~\ref{fig:DQD}(b)] does not involve the
electron-phonon scattering.
The electron only tunnels from the left electronic reservoir to the right one, and vice versa. The elastic currents can be obtained through the Landauer-Buttiker formula~\cite{datta}
\begin{equation}~\label{ie}
\begin{aligned}
I_e^{L,\rm el} &= \int \frac{dE}{2\pi}{\mathcal T(E)}[f_{L}(E) - f_R(E)],\\
I_Q^{L ,\rm el} &= \int \frac{dE}{2\pi}(E-\mu_L){\mathcal T(E)}[f_L(E) - f_R(E)],\\
I_Q^{\rm ph,\rm el} &= 0,
\end{aligned}
\end{equation}
where ${\mathcal T(E)}$ is the transmission function and  can be obtained from the Caroli formula~\cite{gchen2005book}.
Note that elastic processes do not contribute to the heat current $I_Q^{\rm ph}$.
For simplicity we  assume that tunneling rates $\gamma_L=\gamma_R$  and energy independent in this work. Both elastic and inelastic processes contribute to nonequilibrium currents, i.e., $I=I^{\rm el}+I^{\rm inel}$.

Then the power done by the reservoirs for the device is expressed as~\cite{OraPRB2010,Ora2015}
\begin{equation}
{\dot W} = I_p^R\Delta\mu,
\end{equation}
where $\Delta\mu=\mu_R-\mu_L\equiv A_p^R$ is the chemical potential difference.
By default, we will consider the temperatures as
$T_R<T_L<T_{\rm ph}$,
where the temperature of the left reservoir is chosen as the reference temperature.
Hence, it should be easier to characterize which task the system is performing.

\subsection{The efficiency of power production, cooling, and heating by performing a single task}
In this part, we introduce the exergy efficiencies for three thermodynamic operations: heat engines, refrigerators, and heat pumps.
The exergy efficiencies is a dimensionless ratio where the device's output exergy diving the input one.  For example, heat engines converts temperature gradients to extract useful work, e.g., thermoelectric engine~\cite{Esposito09}. In such a thermoelectric heat engine, the heat current absorbed from the hot reservoir (phonon bath) is used to drive particle current against the chemical potential bias $\Delta\mu$, i.e., $I_Q^R<0$, $I_Q^{\rm ph}>0$, and $\dot W<0$. According to the Eq.~\eqref{eq:phi},  the efficiency of the thermoelectric heat engine is given by
\begin{equation}
\phi_{\rm E} = \frac{-{\dot W}}{I_Q^R A_Q^R + I_Q^{\rm ph}A_Q^{\rm ph}}.
\label{eq:phiE}
\end{equation}
Based on the above relation ($\phi_{\rm E}>0$), we know that the negative entropy production associated with power done by the reservoirs, ${\dot W}<0$, is compensated by the positive entropy production of $I_Q^R A_Q^R + I_Q^{\rm ph}A_Q^{\rm ph}$, in agreement with Kedem and Caplan~\cite{Kedem}.

While for the refrigerator operation, it appears when heat current flows out of the coldest reservoir ($I_Q^R>0$), but no other useful task is performed (i.e., $I_Q^{\rm ph}>0$ and $\dot W>0$).
Such efficiency is also termed as the coefficient of performance, defined as
\begin{equation}
\phi_{\rm R} = \frac{-I_Q^R A_Q^R}{I_Q^{\rm ph}A_Q^{\rm ph} + {\dot W}}.
\label{eq:phiR}
\end{equation}

The third type of thermal operations is the heat pump, which is characterized as heat current flowing into the hot reservoir ($I_Q^{\rm ph}<0$), but no other useful task being performed ($I_Q^R<0$ and $\dot W>0$).
For the heat pump, we obtain~\cite{BijayPRL21},
\begin{equation}
\phi_{\rm P} = \frac{-I_Q^{\rm ph}A_Q^{\rm ph}}{I_Q^R A_Q^R + {\dot W}}.
\end{equation}
From the above discussion, we know that if we ignore the inelastic scattering processes, the heat current involved with the phonon bath becomes vanishing.
The whole entropy production is simplified to $T_L\frac{dS^{\rm el}_{\rm tot}}{dt} = I_p^{R,\rm el}  A_p^R  +  I_Q^{R,{\rm el}}  A_Q^R$.
And the three-terminal setup is reduced to the two-terminal one.
Hence, it is obvious to demonstrate that the thermal machines can only perform single task.
In other words, only when multiple terminals cooperate with each other, the thermal machine may perform  multiple useful tasks simultaneously, e.g., the third terminal is either a phononic or an electronic reservoir (see Appendix~\ref{sec1QD}).

\begin{figure}[htb]
\includegraphics[height=9.5cm]{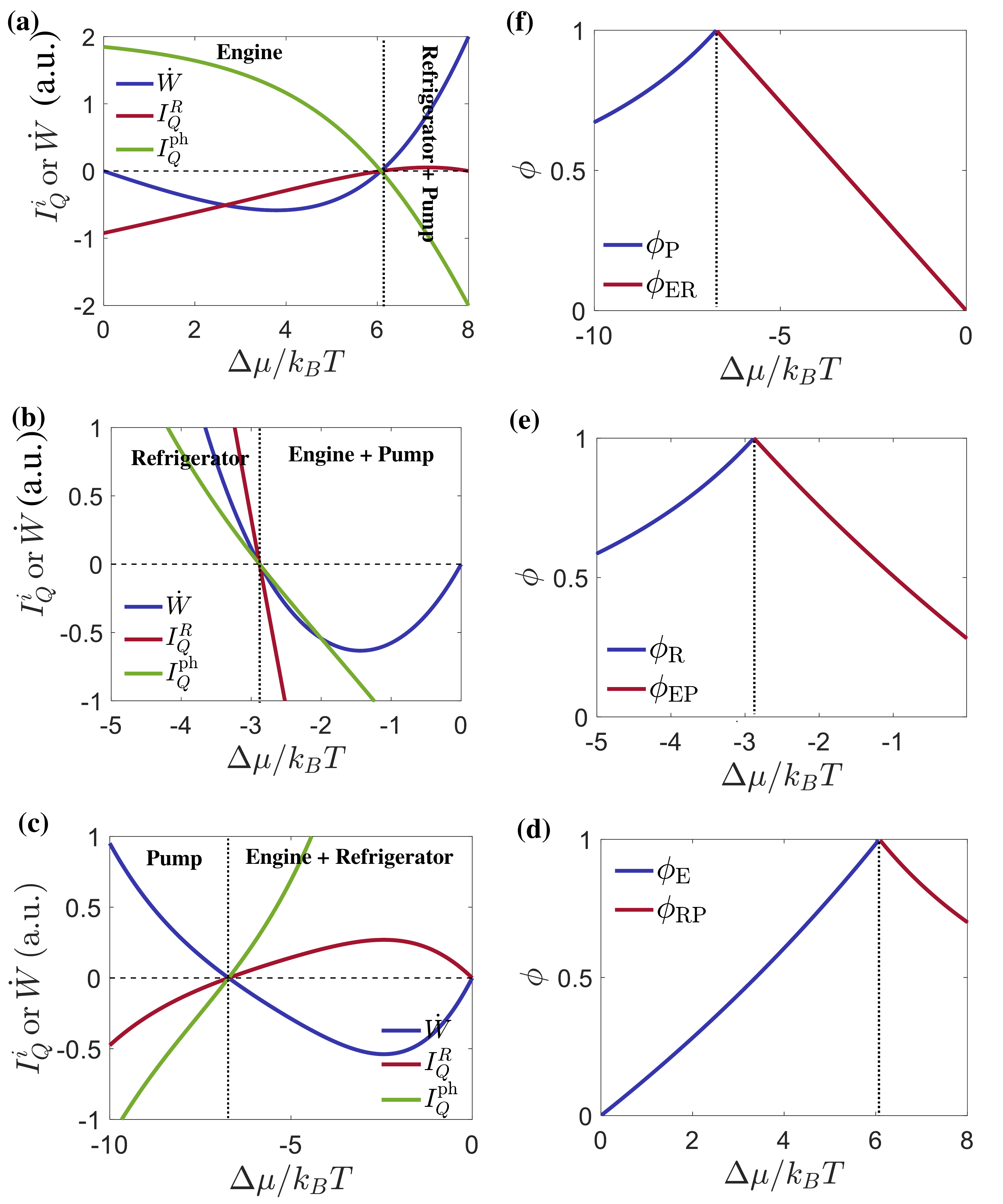}
\caption{Three-terminal double QD system performance. (a)-(c) Currents and power, and (d)-(f) efficiencies, as a function of the chemical potential difference $\Delta\mu$ for simultaneous heating, cooling, and power production, where (a) and (d) $E_L=4 k_BT$, $E_R=-4k_BT$, (b) and (e) $E_L=12 k_BT$, $E_R=10k_BT$, (c) and (f) $E_L=-12 k_BT$, $E_R=0$. The other parameters are $t=0.15 k_BT$, $\gamma_l=\gamma_r=0.4k_BT$, $\gamma_{\rm e-ph}=0.1k_BT$, $k_BT_R=k_BT$, $k_BT_{\rm ph}=5.0k_BT$, $k_BT_L=1.5k_BT$ and $k_BT=10\, {\rm meV}$.}
\label{fig:DQD_IQ}
\end{figure}

\subsection{The performance of the hybrid thermal machine for multiple tasks}
We will show that the three-terminal configuration is  sufficient to perform nontrivial multiple thermodynamic tasks~\cite{ManzanoPRR}. First, work can be used to cool down the cold reservoir and warm up the hot reservoir (i.e., $I_Q^R>0$, $I_Q^{\rm ph}<0$, and $\dot W>0$).
Hence, the device simultaneously acts as a refrigerator and a heat pump.
Specifically, using Eq.~\eqref{eq:phi}, the efficiency in this operation regime reads
\begin{equation}
\phi_{\rm PR} = \frac{-I_Q^R A_Q^R - I_Q^{\rm ph}A_Q^{\rm ph}} {\dot W},
\end{equation}
where the subscript stands for {\it pump-refrigerator}.
And the efficiency for producing work and cooling  the cold reservoir (i.e., $I_Q^R>0$, $I_Q^{\rm ph}>0$, and $\dot W<0$) is given by
\begin{equation}
\phi_{\rm ER} = \frac{-I_Q^R A_Q^R - \dot W} {I_Q^{\rm ph}A_Q^{\rm ph}}.
\end{equation}
While for heat engine and heat pump (i.e., $I_Q^R<0$, $I_Q^{\rm ph}<0$, and $\dot W<0$),
the corresponding efficiency is expressed as
\begin{equation}
\phi_{\rm EP}  = \frac{-I_Q^{\rm ph}A_Q^{\rm ph} - \dot W} {I_Q^R A_Q^R}.
\label{eq:phiEP}
\end{equation}
From the definitions in Eqs.\eqref{eq:phiE}-\eqref{eq:phiEP},
it is clear to find  that the exergy efficiency better characterizes the function of thermal machine, compared with the convention thermodynamic efficiency~\cite{MyJAP,CPB}.
Moreover, we should point out the system is unable to simultaneously heat the hottest reservoir, cool the coldest reservoir, and produce work, which stems from the constraint by the second law of thermodynamics, i.e., $\phi \le 1$~\cite{PRXQuantum}.

The versatility of this setup is manifested in Fig.~\ref{fig:DQD_IQ}.
In particular, we consider the possibility of reproducing the hybrid configurations, e.g., engine-pump, engine-refrigerator, and refrigerator-pump.
We include two separate contributions associated with two different tasks being performed simultaneously. Here we show how  heat currents/output work [Fig.~\ref{fig:DQD_IQ}(a)-\ref{fig:DQD_IQ}(c)] as well as the corresponding efficiencies
[Fig.~\ref{fig:DQD_IQ}(d)-\ref{fig:DQD_IQ}(f)] efficiently characterize thermodynamic multitask, with respect to the chemical potential difference $\Delta\mu$~\cite{HolubecPRL}. Through varying $\Delta\mu$, we find the operation switch between a hybrid regime and the complementary single-task regime. Based on Eqs.~\eqref{eq:currents} and \eqref{ie}, the corresponding vanishing position for currents, termed the current cutoff voltage, can be obtain as $I_p^R\approx 0$. For example, in Fig.~\ref{fig:DQD_IQ}(d), we see that when $0<\Delta\mu < 6 k_BT$, the device produces powers to the reservoirs, which performs the task of being a heat engine ${\dot W}>0$ (see the blue line).
As $\Delta\mu > 6 k_BT$, the device will heat the hottest reservoir $I_Q^{\rm ph}<0$ and cool the coldest reservoir $I_Q^R>0$, which acts as hybrid thermal machines.
In analogy,  we see a similar behavior in Figs.~\ref{fig:DQD_IQ}(e) and \ref{fig:DQD_IQ}(f).
When $\Delta\mu \approx -3 k_BT$ ($-7 k_BT$), the device will change from a single task of the refrigerator (heat pump) to two simultaneous tasks of heat engine and heat pump (engine and refrigerator).

We also note that thermodynamic laws set a bound on the efficiency of the quantum thermal machine,
where the output work and efficiency cannot be maximized simultaneously. Such restriction refers to the power-efficiency trade-off~\cite{Van2005,JiangPRE,PED,BrandnerPRX,Naoto,Constancy,trade-off}, e.g., the operating efficiency reaching the unit and the output work vanishing.
In addition to the above functions, our three-terminal setups characterized as Eq.~\eqref{eq:phiR} can also be designed for cooling by heating processes.
As proposed by Cleuren {\it et al}.\cite{Cooling2}, the chemical potential difference is kept zero and the coolest electronic reservoir is cooled by the hot phonon bath.

\subsection{Linear transport and thermoelectric cooperative effect}
Until now, we focus on the thermoelectric energy conversion efficiency of the three-terminal quantum thermal machine, of which two main parameters for characterizing thermodynamic performance are the figure of merit $\xi$ and power factor $P$.
 Next, we demonstrate in detail how the cooperative effect can improve the exergy efficiency (figure of merit) in the linear response regime.

Based on Eqs.~\eqref{eq:currents} and \eqref{ie}, the transport equations are reexpressed as~\cite{Jiang2012,Jiang2013,JiangJAP}
\begin{equation}
\begin{aligned}
\left( \begin{array}{cccc} I_p^R\\ I_Q^R \\ I_Q^{\rm ph} \end{array}\right) =
\left( \begin{array}{cccc} M_{11} & M_{12} & M_{13} \\ M_{12} & M_{22} & M_{23} \\ M_{13} & M_{23} & M_{33}
        \end{array} \right)
\left( \begin{array}{cccc} A_p^R \\  A_Q^R \\ A_Q^{\rm ph}
        \end{array}\right),
\label{eq:Onsager}
\end{aligned}
\end{equation}
where the Onsager coefficients $M_{ij}=M_{ij}^{\rm el}+M_{ij}^{\rm inel}$ include both the elastic and inelastic transport components~\cite{Jiangtransistors}.
Moreover, the second law of thermodynamics, i.e., $dS/dt\ge0$~\cite{datta}, leads to
\begin{equation}
\begin{aligned}
M_{11},M_{22},M_{33}\ge0, \quad  M_{11}M_{22}\ge M_{12}^2, \\
M_{11}M_{33}\ge M_{13}^2, \quad M_{22}M_{33}\ge M_{23}^2.
\end{aligned}
\end{equation}
And the determinant of the transport matrix in Eq.~\eqref{eq:Onsager} should be non-negative.

For systems with time-reversal symmetry, the maximal efficiency and maximal power of the three-terminal devices are expressed as
\begin{equation}
\phi_{\max} = \frac{\sqrt{\xi +1} - 1}{\sqrt{\xi + 1} + 1}, \quad W_{\max}=\frac{1}{4}PT^2,
\end{equation}
where $\xi$ is the dimensionless figure of merit and $P$ is the power factor, respectively~\cite{WhitneySciPost}. The maximal efficiency quantifies the performance of hybrid thermal machines, including heat engines, refrigerators, heat pumps and their combinations. Clearly, $\phi_{\max}$ approaches the unit when $\xi$ approaches $\infty$.

Alternatively from the geometric perspective, the two temperatures can be parametrized as~\cite{JiangJAP,MyJAP,Niedenzu18NJP,CPB,CooperativeSpin} $A_Q^R = T_A \cos\theta$, and $A_Q^{\rm ph} = T_A \sin\theta$. Consequently, the figure of merit and power factor are described as

\begin{equation}
\frac{1}{\xi (\theta)}= \frac{M_{22}\cos^2\theta+2M_{23}\sin\theta\cos\theta+M_{33}\sin^2\theta}{M_{11} (S_1\cos\theta + S_2\sin\theta)^2}-1,
\label{zzt}
\end{equation}

\begin{equation}
P(\theta) = M_{11}(S_1\cos\theta + S_2\sin\theta)^2 .
\label{ppower}
\end{equation}
where $S_1=M_{12}/M_{11}$ and $S_2=M_{13}/M_{11}$  denote the longitudinal and transverse thermopowers, respectively. Using Eqs.~(\ref{zzt}) and (\ref{ppower}),  it is straightforward to obtain the expressions of the longitudinal ($\theta=0$ or $\pi$) and transverse  ($\theta=\pi/2,3\pi/2$) figure of merit and power factor.
The details can be found in Table \ref{tablongtran}, where ${\mathcal M}=M_{11} M_{22}M_{33} - M_{11} M_{23}^2 -  M_{33} M_{12}^2 + 2 M_{12}M_{13}M_{23}- M_{22} M_{13}^2$.
Specifically, we find that $\xi_{\max}  \ge \max(\xi_L,\xi_T)$ and $P_{\max}\ge \max({P_L,P_T})$.

\begin{figure}[htb]
\includegraphics[height=4.0cm]{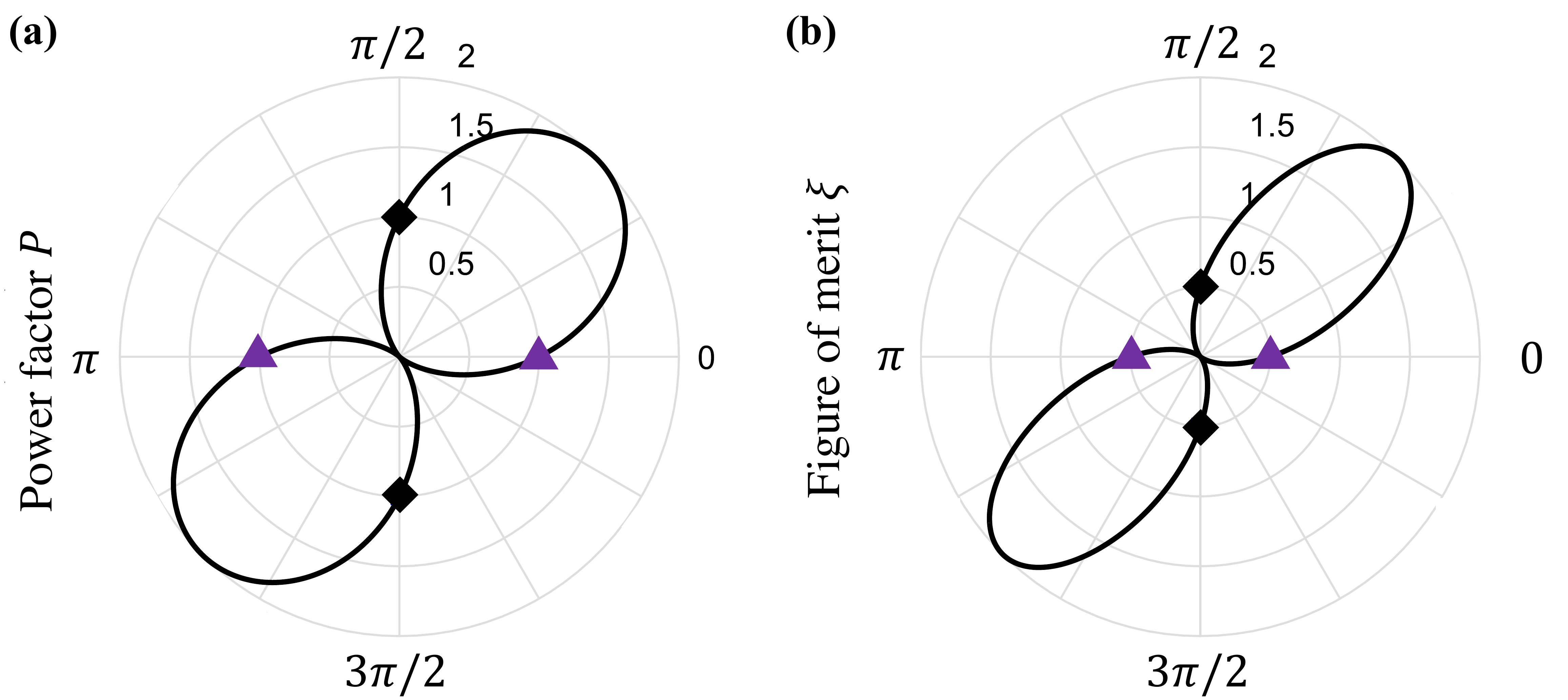}
\caption{Polar plot of (a) power factor $P(\theta)$ and (b) figure of merit $\xi(\theta)$ [in arbitrary unit (a.u.)] versus angle $\theta$. At $\theta=0$ or $\pi$, $\xi$ and $P$ recover the values for the longitudinal thermoelectric effect (purple triangles, $\xi_L$ and $P_L$), while at $\theta=\pi/2$ and $3\pi/2$, they recover the transverse thermoelectric effect (black rhombus, $\xi_T$ and $P_T$). The parameters are same with Fig.~\ref{fig:DQD_IQ}.}
\label{fig:ZTPtheta}
\end{figure}

\begin{widetext}

\begin{table}[htb]
\caption{Longitudinal and transverse thermoelectric effects and their maximum}
\setlength{\tabcolsep}{3mm}
\begin{tabular}{llllllllll}\hline 
Peformance   & \mbox{} & Figure of merit & \mbox{} & Power factor   \\
\hline
Longitudinal & \mbox{} & $\xi_L = \frac{M_{11} S_1^2}{M_{22} - M_{11}S_1^2 }$  &   \mbox{} &  $P_{L} = M_{11}S_1^2$ \\
Transverse    & \mbox{} & $ \xi_T = \frac{M_{11} S_2^2}{M_{33} - M_{11} S_2^2 }$  & \mbox{} & $P_{T} = M_{11} S_2^2$ \\
Maximum  & \mbox{} &  $\xi_{\max} = \frac{M_{33} M_{12}^2 - 2 M_{12}M_{13}M_{23}+ M_{22} M_{13}^2}{\mathcal M}$  & \mbox{} & $P_{\max} = M_{11}(S_1^2+S_2^2)$ \\
\hline
\end{tabular}
\label{tablongtran}
\end{table}

\end{widetext}

Figure \ref{fig:ZTPtheta} shows the figure of merit $\xi$ and power factor $P$ by tuning the angle $\theta$ in a polar plot. We demonstrate the robustness of the cooperative effects by examining the renormalized factor of power $P_{\max}/\max(P_L,P_T)$ and the renormalized factor of  efficiency $\xi_{\max}/\max(\xi_L,\xi_T)$. Outstandingly, the maximum figure of merit $\xi_{\rm max}$ can be greater than the longitudinal figure of merit $\xi_L$ (the purple triangle in Fig.~\ref{fig:ZTPtheta}(b)) and transverse one $\xi_T$ (the black rhombus in Fig.~\ref{fig:ZTPtheta}(b)) for $0<\theta<\pi/2$ and $\pi<\theta<3\pi/2$.
In order to further understand the physical mechanism of thermoelectric cooperation effect,
we know from Eq.~\eqref{eq:Onsager} that the particle current is expressed as $I_p^R=M_{11}A_p^R+M_{12}A_Q^R+M_{13}A_Q^{\rm ph}$, where $M_{12}A_Q^R$ represents the longitudinal thermoelectric effect and $M_{13}A_Q^{\rm ph}$ represents the transverse one.
 As $0<\theta<\pi/2$ and $\pi<\theta<3\pi/2$,
 it is found that $(M_{12}A_Q^R){\times}(M_{13}A_Q^{\rm ph})>0$
 and the longitudinal and transverse thermoelectric effects are coherently superposed.
 The similar physical effect can be found  for the power factor.
 Therefore, we conclude that the thermoelectric figure of merit and power factor can be enhanced by exploiting cooperative effects.

\begin{figure}[htb]
\begin{center}
\centering\includegraphics[width=8.9cm]{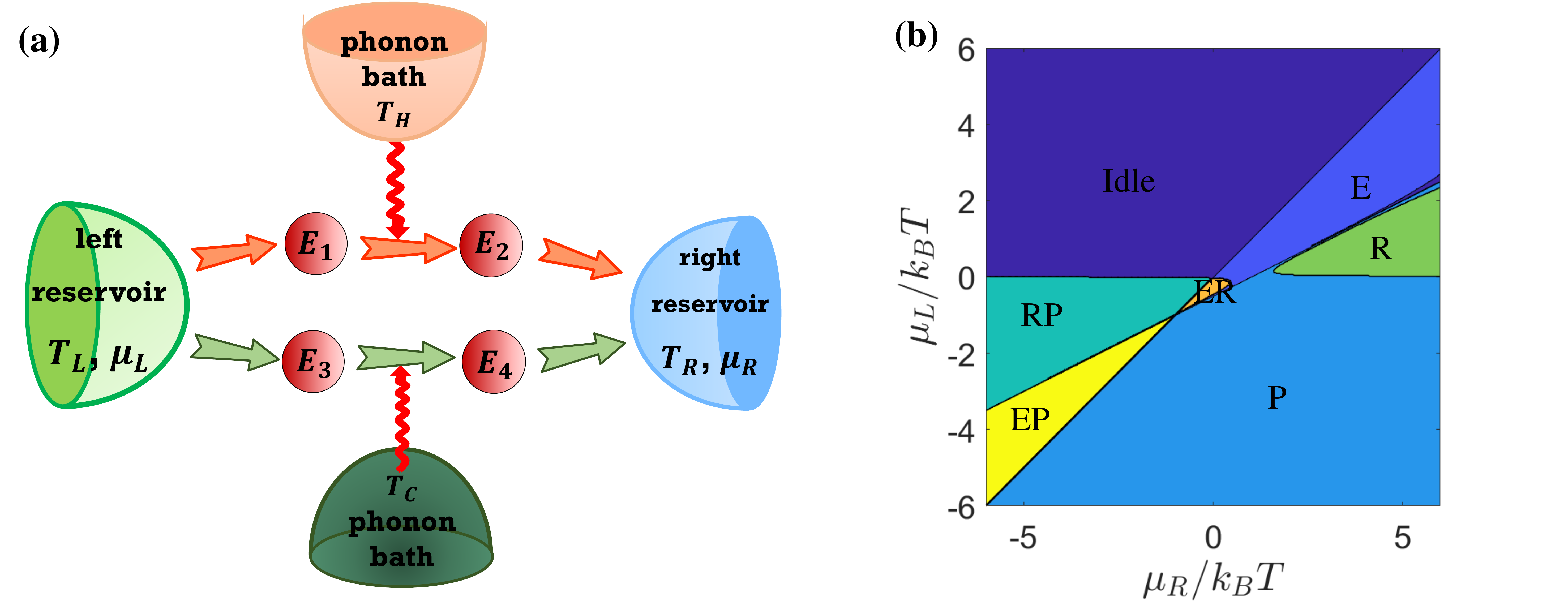}
\caption{(a) Schematic of four-terminal four-QD thermoelectric systems. The left reservoir and right reservoir with chemical potentials $\mu_{L/R}$ and temperatures $T_{L/R}$ connects with the central QD systems, in which there are two parallel transport channels: the upper channel has two QDs (with energies $E_1$ and $E_2$) and a phonon bath $H$ (with temperature $T_H$); the lower channel has two QDs (with energies $E_3$ and $E_4$) and a phonon bath $C$ (with temperature $T_C$). The two channels are spatially separated so that the phonon bath $H$ ($C$) couples only to the upper (lower) channel. (b) Map of the system functionality as the functions of the chemical potentials $\mu_L$ and $\mu_R$. The parameters are: $E_1=2.0k_BT$, $E_2=0$, $E_3=1.0k_BT$, $E_4=5.0k_BT$, $\gamma_{\rm e-ph}=0.1k_BT$, $k_BT_L=k_BT$, $k_BT_C=0.7k_BT$, $k_BT_H=2.0k_BT$, $k_BT_R=0.5k_BT$, and $k_BT=10\, {\rm meV}$.} ~\label{fig:4T-model}
\end{center}
\end{figure}

\section{The four-terminal thermoelectric QD systems}~\label{sec4QD}

To further illustrate the functionality of inelastic hybrid thermal machines, we consider another typical thermoelectric transport model, i.e., four-terminal thermoelectric QD systems. In Ref.~\cite{MyPRBdemon}, we studied a unconventional inelastic thermoelectric effect, termed as cooling by transverse heat current effect.
It describes the cooling process of the source driven by the heat exchange between the two thermal baths, rather than total heat injected into the central quantum system~\cite{Cooling2}.
In this section, we study how to use the temperature gradient to overcome the chemical potential different to generate useful work (heat engines), or use the chemical potential to against the temperature bias
to generate the heat current  from the hot reservoir to cold one (refrigerators and/or heat pumps).
Interestingly, this device  performs multiple useful tasks simultaneously, which demonstrates that more thermodynamic forces enrich the implementation of hybrid operations.


The mesoscopic four-terminal thermoelectric devices coupled with two electric reservoirs and two phononic reservoirs as depicted in Fig.~\ref{fig:4T-model}(a).
The system consists of four QDs: QD 1 and QD 2 with the energies $E_1$ and $E_2$ in the upper channel are coupled with the hot phononic reservoir $H$, while QD 3 and QD 4 in the lower channel with energy $E_3$ and $E_4$ are connected with the cold phononic bath $C$. The inelastic-scattering processes dominate the thermoelectric transport~\cite{WhitneyPhysE,MyCPL21}. $I_p^i$ ($i=L,R$) is the particle current flowing from the reservoir $i$ into system and $I_Q^i$ ($i=L,R,H,C$) is the heat current flows from the reservoir $i$ into the system.

Due to energy conservation ($I_Q^L + {\mu_L}I_p^L + I_Q^H + I_Q^C + I_Q^R + {\mu_R}I_p^R=0$) and particle conversation ($I_p^L+I_p^R=0$),  the entropy production rate of the whole system is given by \cite{MyPRBdemon}
\begin{equation}
T_L\frac{dS}{dt} = I_p^R X_p^R + I_Q^R X_Q^R + I_Q^{H}X_Q^H + I_Q^C X_Q^C,
\end{equation}
where the affinities are defined as $X_p^R = \mu_R - \mu_L$, $X_Q^R = 1 - \frac{T_L}{T_R}$, $X_Q^C = 1 - \frac{T_L}{T_C}$, $X_Q^H = 1 - \frac{T_L}{T_H}$. Here, we restrict our discussion where there is only one energy level in each QD.
In this setup, the phonon-assisting  particle currents through  two independent upper and lower channels are described as~\cite{MyPRBtransistor},
\begin{equation}
I_{\rm up} = \Gamma_{1\rightarrow2} - \Gamma_{2\rightarrow1} ,\quad
I_{\rm down} = \Gamma_{3\rightarrow4} - \Gamma_{4\rightarrow3}
\label{eq:I1234}
\end{equation}
where $\Gamma_{i\rightarrow j} $ is the electron tunneling rate from QD $i$ to QD $j$~\cite{MyPRBdemon}. Then, the particle and heat currents derived from the Fermi golden rule~\cite{Jiangtransistors} can be written as
\begin{equation}
\begin{aligned}
I_p^L & = I_{\rm up} + I_{\rm down},  \,\quad  I_Q^L = (E_1-\mu_L)I_{\rm up} + (E_3-\mu_L)I_{\rm down}, \\
I_Q^R & = -(E_2-\mu_R)I_{\rm up} - (E_4-\mu_R)I_{\rm down}, \\
I_Q^H & = (E_2-E_1)I_{\rm up}, \,\quad   I_Q^C = (E_4-E_3)I_{\rm down}.
\label{eq:4TIA}
\end{aligned}
\end{equation}
And the output power of the device $-{\dot W}$ is given by
\begin{equation}
{\dot W} = I_p^R\Delta\mu,
\end{equation}
where $\Delta\mu = \mu_R-\mu_L\equiv X_p^R$ is the chemical potential difference. In this four-terminal systems, we set the temperatures as
\begin{equation}
T_R<T_C< T_L<T_H.
\end{equation}
The definitions of efficiency for the four-terminal device in different operation regimes are shown in Table \ref{table4QD}.
In our setup, heating the hottest reservoir and cooling the coolest reservoir are considered as useful processes, and this part of entropy production is reduced.
In contrast, heating or cooling an intermediate temperature heat reservoir is considered as a useless process.
In the situation with $X_p^R=0$ and $T_C=(2/T_L-1/T_H)^{-1}$, the efficiency of refrigerators is simplified as $\phi=-(2I_Q^R)/(I_Q^H-I_Q^C)$.
It is convenient to discover a mode of cooling by transverse heat current effect: the coolest right reservoir can be cooled by passing a heat current between the $H$ phonon bath and the $C$ phonon bath, with no energy exchange between the phonon and electronic reservoirs~\cite{MyPRBdemon,MyCPL21}. Obviously, the cooling by thermal current operation requires a higher temperature of phonon bath than the cooling by electric power.
Hence, we mainly consider the latter situation when the temperature gradient is comparatively small.

\begin{widetext}

\begin{table}[htb]
\caption{Functionality of four-terminal four-QD thermal machine}
\setlength{\tabcolsep}{3mm}
\begin{tabular}{llllllllll}\hline 
Thermal progresses   & \mbox{} & work and heat currents & \mbox{} &Efficiency   \\
\hline
Heat engine & \mbox{} & $I_Q^R<0$, $I_Q^H>0$, $\dot W<0$, $I_Q^C<0$  &   \mbox{} &  $\phi_{\rm E} = \frac{-{\dot W}}{I_Q^R X_Q^R + I_Q^C X_Q^C + I_Q^H X_Q^H}$ \\
Heat pump  & \mbox{} &  $I_Q^H<0$, $I_Q^R<0$, $\dot W>0$, $I_Q^C<0$  & \mbox{} & $\phi_{\rm P} = \frac{-I_Q^H X_Q^H}{I_Q^R X_Q^R +I_Q^C X_Q^C +  {\dot W}}$ \\
Refrigerator & \mbox{} & $I_Q^R>0$, $I_Q^H>0$, $\dot W>0$, $I_Q^C<0$ &  \mbox{} & $\phi_{\rm R} = \frac{-I_Q^R X_Q^R}{I_Q^H X_Q^H +I_Q^C X_Q^C +  {\dot W}}$  \\
Refrigerator and heat pump & \mbox{} & $I_Q^R>0$, $I_Q^H<0$, $\dot W>0$, $I_Q^C<0$  &  \mbox{} & $\phi_{\rm PR} = \frac{-I_Q^R X_Q^R - I_Q^H X_Q^H} {I_Q^C X_Q^C + \dot W}$ \\
Engine and refrigerator & \mbox{} & $I_Q^R>0$, $I_Q^H>0$, $\dot W<0$, $I_Q^C<0$   &  \mbox{} & $\phi_{\rm ER} = \frac{-I_Q^R X_Q^R - \dot W} {I_Q^C X_Q^C + I_Q^H X_Q^H}$ \\
Engine and heat pump & \mbox{} &   $I_Q^R<0$, $I_Q^H<0$, $\dot W<0$, $I_Q^C<0$  & \mbox{} & $\phi_{\rm EP} = \frac{-I_Q^H X_Q^H - \dot W} {I_Q^C X_Q^C + I_Q^R X_Q^R}$ \\
\hline 
\end{tabular}
\label{table4QD}
\end{table}

\end{widetext}

Fig.~\ref{fig:4T-model}(b) displays all possible functionalities for the thermal machines, and the change of current symbol means the realization of different functions, by tuning the chemical potential of left ($\mu_L$) and right ($\mu_R$) reservoirs. From this map figure, we observe that the four-terminal device can realize any functions and perform different tasks by adjusting physical parameters.

\begin{widetext}

\begin{figure}[htb]
\includegraphics[height=8.0cm]{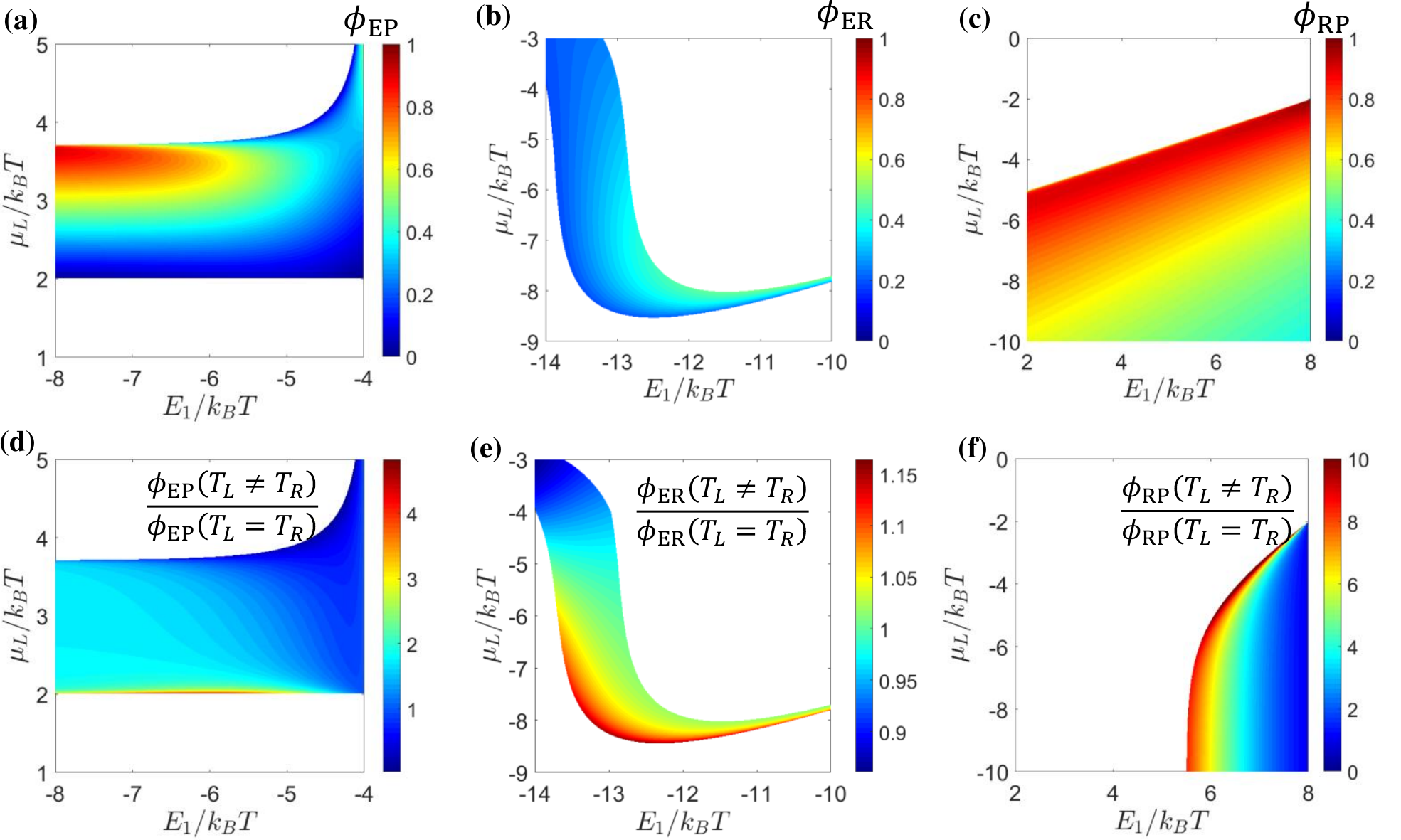}
\caption{The efficiencies of hybrid thermal machines (a) $\phi_{\rm EP}$, (b) $\phi_{\rm ER}$ and (c) $\phi_{\rm RP}$ as the functions of the chemical potential $\mu_L$ and QD energy $E_1$ for $T_C=0.7k_BT$. (d)-(f) Comparing the efficiency for $T_L\ne T_C$ case and the $T_L=T_C$ as the functions of the chemical potential $\mu_L$ and QD energy $E_1$. The parameters: (a) and (d) $E_2=-4.0 k_BT$, $E_3=2.0k_BT$, $E_4=-1.0k_BT$. (b) and (e) $E_2=6.0 k_BT$, $E_3=6.0k_BT$, $E_4=8.0k_BT$. (c) and (f) $E_2=8.0k_BT$, $E_3=6.0k_BT$, $E_4=-4.0k_BT$. The other parameters are $\gamma_{\rm e-ph}=0.1k_BT$, $k_BT_L=k_BT$, $k_BT_H=2.0k_BT$, $k_BT_R=0.5k_BT$, and $k_BT=10\, {\rm meV}$. The white region indicates the function of the hybrid thermal machines is not performed.}
\label{fig:4QDefficiency}
\end{figure}

\end{widetext}

In order to emphasize the role of thermodynamic biases in the four-terminal system, we compare the performance of such setup at $T_L=T_R$ with the case $T_L\ne T_R$ for the same three-terminal quantum-dot thermal machines in Fig.~\ref{fig:4QDefficiency}. We first show the efficiency of the thermal machines in Figs.~\ref{fig:4QDefficiency}(a)-\ref{fig:4QDefficiency}(c), which exhibits two useful tasks as functions of QD energy $E_1$ and chemical potential $\mu_L$, and the efficiency can even reach the unity in some parameter regions. In Fig.~\ref{fig:4QDefficiency}(d)-\ref{fig:4QDefficiency}(f), we compare the above two cases and it is easy to find that the maximum efficiency can be significantly improved when the four-terminal device goes from the condition of $T_L=T_R$  to $T_L\ne T_R$. Through concrete numerical results, we find that the setup with multiple biases can substantially enlarge the parameter region of the high efficiency, and thus provide a promising pathway toward high-performance thermal machines.

Finally, we point out that there are three Seebeck coefficients induced by three temperature gradients, which originate from the three thermal affinities, $X_Q^R$, $X_Q^H$, and $X_Q^C$ in the linear-response regime for the four-terminal thermoelectric device. Correspondingly, the figures of merit refer to the conventional longitudinal thermoelectric effects and the unconventional transverse thermoelectric effects, respectively. The high figure of merit for unconventional thermoelectric effects require the small variance of the phonon energy, but the electron energy variance.

\section{Conclusions}~\label{conclusion}

In this work, a general thermodynamic efficiency, termed as ``exergy efficiency", was defined from an entropic point of view,
which can characterize the functions of the thermal machines, such as heat engine, refrigerator, and heat pump, without the need to specifically define a reference temperature. The exergy efficiency is appointed for each functionality as the ratio of the output (the consumed usable energy) to the consumed (the target heat).

We discussed  thermodynamic operations of phonon-thermoelectric device that perform multiple useful tasks simultaneously in two typical thermoelectric systems [(i) devices with two electronic and a phonon terminals, (ii) those with two electronic terminals and two phononic terminals], where phonon-assisted inelastic processes dominate the transport. In the three-terminal device, we found the parameter region and the corresponding exergy efficiency of the thermal machines for power production, cooling, heating, and the simultaneous combinations.
It is found that a hybrid thermal machine can be realized only if inelastic processes are taken into account. While in the four-terminal device, in addition to demonstrating that the device can perform two useful tasks, we further emphasized the improvement of exergy efficiency by multiple thermodynamic biases.

Moreover, in the linear-response regime, from the geometric aspect we demonstrated that the thermoelectric cooperative effects can enhance the performance of multitasks, e.g., efficiency and output power, which are tightly related with the figure of merit and power factor. Due to the inelastic thermoelectric effect, particle and heat currents can flow in spatially separated parts of the multiterminal systems. Our analytical results revealed the importance of inelastic transport effects for the design of the high performance thermoelectric device.

Finally, it should be pointed out that our study is based on the steady-state transport. The performance of the periodically driven thermoelectric device with inelastic transport undertaking multitasks is fascinating in future study.

\section{ACKNOWLEDGEMENTS}
We are grateful to Professor Jie Ren for many interesting discussions. This work was supported by the funding for the National Natural Science Foundation of China under Grants No. 12125504, No. 12074281, and No. 11704093, the Opening Project of Shanghai Key Laboratory of Special Artificial Microstructure Materials and Technology, and Jiangsu Key Disciplines of the Fourteenth Five-Year Plan (Grant No. 2021135).

\appendix

\section{The hybrid thermal machines: three-terminal single-level QD System}\label{sec1QD}
\begin{figure}
\begin{center}
\centering \includegraphics[width=8.5cm]{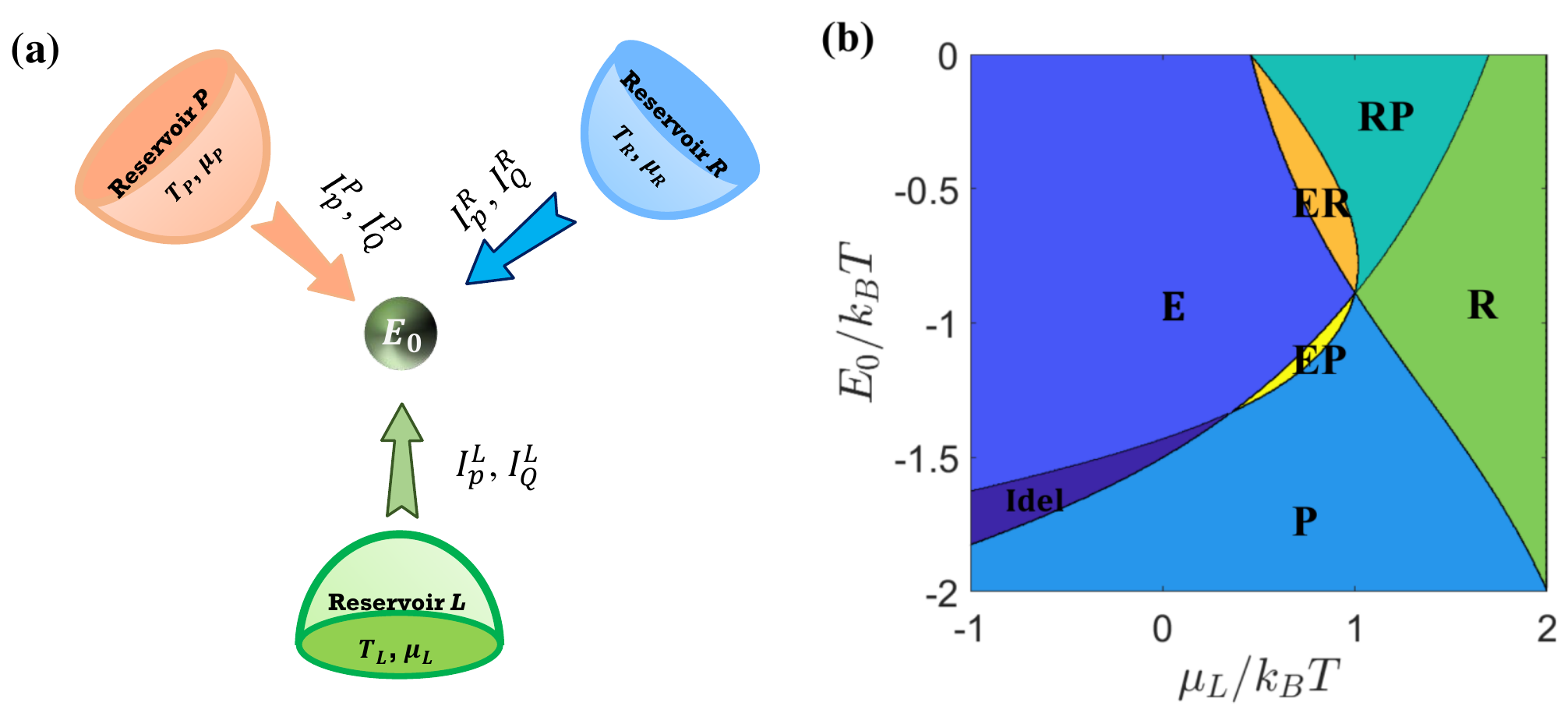}
\caption{(a) Schematic figure of a three-terminal single-QD thermoelectric device. The QD with a single energy level $E_0$ are connected in series to three fermionic reservoirs $i$ ($i=L,R,P$). The chemical potential and temperature of reservoirs are $\mu_i$ and $T_i$. The constant $\Gamma$ represents the coupling between the QD and reservoir. (b) Map of the system efficiency as the functions of the chemical potential $\mu_L$ and QD energy $E_0$. The parameters are $\mu_R=0$, $\mu_P=2.0k_BT$.}
\label{figure:3T-1QD}
\end{center}
\end{figure}

In this appendix,  we  study the operation and performance of the elastic thermoelectric devices that perform multiple useful tasks simultaneously. In our construction (see Fig.~\ref{figure:3T-1QD})(a), a single QD system exchanges particle and energy with three electronic reservoirs, $L$, $R$, and $P$. The particle and heat currents flowing from the reservoir into the system are expressed as\cite{gchen2005book}
\begin{subequations}
\begin{align}
&I_p^L=\sum_{i=R,P}\frac{\Gamma_i\Gamma_L[f_L(E_0) - f_i(E_0)]}{\Gamma_L+\Gamma_i},\\
&I_p^R=\sum_{i=L,P}\frac{\Gamma_i\Gamma_R[f_R(E_0) - f_i(E_0)]}{\Gamma_R+\Gamma_i},\\
&I_p^P=\sum_{i=L,R}\frac{\Gamma_i\Gamma_P[f_P(E_0) - f_i(E_0)]}{\Gamma_P+\Gamma_i},
\end{align}
\end{subequations}
and
\begin{subequations}
\begin{align}
&I_Q^L=\sum_{i=R,P}\frac{(E_0-\mu_L)\Gamma_i\Gamma_L[f_L(E_0) - f_i(E_0)]}{\Gamma_L+\Gamma_i},\\
&I_Q^R=\sum_{i=L,P}\frac{(E_0-\mu_R)\Gamma_i\Gamma_R[f_R(E_0) - f_i(E_0)]}{\Gamma_R+\Gamma_i},\\
&I_Q^P=\sum_{i=L,R}\frac{(E_0-\mu_P)\Gamma_i\Gamma_R[f_P(E_0) - f_i(E_0)]}{\Gamma_P+\Gamma_i},
\end{align}
\end{subequations}
respectively. Particle conservation implies that $I_p^L+I_p^R+I_p^P=0$, while energy conservation requires ${I_Q^L}+\mu_LI_p^L+{I_Q^R}+\mu_RI_p^R+I_Q^P+\mu_PI_p^P=0$~\cite{JiangCRP}.

Considering the energy and particle conversations, the entropy production is given by a specific form,
\begin{equation}
\begin{aligned}
T_L\frac{dS}{dt}& = F_p^RI_p^R + F_p^PI_p^P + F_Q^RI_Q^R + F_Q^PI_Q^P.
\end{aligned}
\end{equation}
where $F_p^R = \mu_R-\mu_L$, $F_p^P = \mu_P-\mu_L$, $F_Q^R = 1 - \frac{T_L}{T_R}$, and $F_Q^P = 1 - \frac{T_L}{T_P}$.

The power done by the reservoirs to the system is given by~\cite{MyJAP}
\begin{equation}
{\dot W} = F_p^RI_p^R + F_p^PI_p^P,
\end{equation}
and in the single QD three-terminal systems, we set
\begin{equation}
T_R<T_L<T_P,
\end{equation}
and choose the temperature of the left reservoir as the reference temperature. The exergy efficiency in six different operational regimes is shown in Table \ref{table1QD}. In Fig.~\ref{figure:3T-1QD}(b), we find the device with three electronic reservoirs also can perform useful multitasks through adjusting the QD energy $E_0$ and chemical potential $\mu_L$.

\begin{widetext}

\begin{table}[htb]
\caption{Functionality of three-terminal single-QD thermal machine}
\setlength{\tabcolsep}{3mm}
\begin{tabular}{llllllllll}\hline 
Thermal progresses   & \mbox{} & work and heat currents & \mbox{} && \mbox{} & Efficiency   \\
\hline
Heat engine & \mbox{} & $I_Q^R<0$, $I_Q^P>0$, $\dot W<0$  &   \mbox{} & & \mbox{} & $\phi_{\rm E} = \frac{-{\dot W}}{I_Q^R F_Q^R + I_Q^PF_Q^P}$ \\
Heat pump  & \mbox{} &  $I_Q^P<0$, $I_Q^R<0$, $\dot W>0$  & \mbox{} && \mbox{} & $\phi_{\rm P} = \frac{-I_Q^P F_Q^P}{I_Q^R F_Q^R + {\dot W}}$ \\
Refrigerator & \mbox{} & $I_Q^R>0$, $I_Q^P>0$, $\dot W>0$ &  \mbox{} & & \mbox{} &$\phi_{\rm R} = \frac{-I_Q^R F_Q^R}{I_Q^PF_Q^P + {\dot W}}$  \\
Refrigerator and pump & \mbox{} & $I_Q^R>0$, $I_Q^P<0$, $\dot W>0$  &  \mbox{} & & \mbox{} &$\phi_{\rm PR} = \frac{-I_Q^R F_Q^R - I_Q^PF_Q^P} {\dot W}$ \\
Engine and refrigerator & \mbox{} & $I_Q^R>0$, $I_Q^P>0$, $\dot W<0$   &  \mbox{} && \mbox{} & $\phi_{\rm ER} = \frac{-I_Q^R F_Q^R - \dot W} {I_Q^PF_Q^P}$ \\
Engine and pump & \mbox{} &   $I_Q^R<0$, $I_Q^P<0$, $\dot W<0$  & \mbox{} & & \mbox{} &$\phi_{\rm EP} = \frac{-I_Q^PF_Q^P - \dot W} {I_Q^R F_Q^P}$ \\
\hline
\end{tabular}
\label{table1QD}
\end{table}

\end{widetext}

\bibliography{Ref-Multitask}

\begin{thebibliography}{81}%
\makeatletter
\providecommand \@ifxundefined [1]{%
 \@ifx{#1\undefined}
}%
\providecommand \@ifnum [1]{%
 \ifnum #1\expandafter \@firstoftwo
 \else \expandafter \@secondoftwo
 \fi
}%
\providecommand \@ifx [1]{%
 \ifx #1\expandafter \@firstoftwo
 \else \expandafter \@secondoftwo
 \fi
}%
\providecommand \natexlab [1]{#1}%
\providecommand \enquote  [1]{``#1''}%
\providecommand \bibnamefont  [1]{#1}%
\providecommand \bibfnamefont [1]{#1}%
\providecommand \citenamefont [1]{#1}%
\providecommand \href@noop [0]{\@secondoftwo}%
\providecommand \href [0]{\begingroup \@sanitize@url \@href}%
\providecommand \@href[1]{\@@startlink{#1}\@@href}%
\providecommand \@@href[1]{\endgroup#1\@@endlink}%
\providecommand \@sanitize@url [0]{\catcode `\\12\catcode `\$12\catcode
  `\&12\catcode `\#12\catcode `\^12\catcode `\_12\catcode `\%12\relax}%
\providecommand \@@startlink[1]{}%
\providecommand \@@endlink[0]{}%
\providecommand \url  [0]{\begingroup\@sanitize@url \@url }%
\providecommand \@url [1]{\endgroup\@href {#1}{\urlprefix }}%
\providecommand \urlprefix  [0]{URL }%
\providecommand \Eprint [0]{\href }%
\providecommand \doibase [0]{http://dx.doi.org/}%
\providecommand \selectlanguage [0]{\@gobble}%
\providecommand \bibinfo  [0]{\@secondoftwo}%
\providecommand \bibfield  [0]{\@secondoftwo}%
\providecommand \translation [1]{[#1]}%
\providecommand \BibitemOpen [0]{}%
\providecommand \bibitemStop [0]{}%
\providecommand \bibitemNoStop [0]{.\EOS\space}%
\providecommand \EOS [0]{\spacefactor3000\relax}%
\providecommand \BibitemShut  [1]{\csname bibitem#1\endcsname}%
\let\auto@bib@innerbib\@empty
\bibitem [{\citenamefont {Sothmann}\ \emph {et~al.}(2015)\citenamefont
  {Sothmann}, \citenamefont {S{\'a}nchez},\ and\ \citenamefont
  {Jordan}}]{Nanotechnology}%
  \BibitemOpen
  \bibfield  {author} {\bibinfo {author} {\bibfnamefont {B.}~\bibnamefont
  {Sothmann}}, \bibinfo {author} {\bibfnamefont {R.}~\bibnamefont
  {S{\'a}nchez}}, \ and\ \bibinfo {author} {\bibfnamefont {A.~N}\ \bibnamefont
  {Jordan}},\ }\bibfield  {title} {\enquote {\bibinfo {title} {Thermoelectric
  energy harvesting with quantum dots},}\ }\href
  {http://stacks.iop.org/0957-4484/26/i=3/a=032001} {\bibfield  {journal}
  {\bibinfo  {journal} {Nanotechnology}\ }\textbf {\bibinfo {volume} {26}},\
  \bibinfo {pages} {032001} (\bibinfo {year} {2015})}\BibitemShut {NoStop}%
\bibitem [{\citenamefont {Jiang}\ and\ \citenamefont {Imry}(2016)}]{JiangCRP}%
  \BibitemOpen
  \bibfield  {author} {\bibinfo {author} {\bibfnamefont {J.-H.}\ \bibnamefont
  {Jiang}}\ and\ \bibinfo {author} {\bibfnamefont {Y.}~\bibnamefont {Imry}},\
  }\bibfield  {title} {\enquote {\bibinfo {title} {Linear and nonlinear
  mesoscopic thermoelectric transport with coupling with heat baths},}\ }\href
  {\doibase https://doi.org/10.1016/j.crhy.2016.08.006} {\bibfield  {journal}
  {\bibinfo  {journal} {C. R. Phys.}\ }\textbf {\bibinfo {volume} {17}},\
  \bibinfo {pages} {1047 -- 1059} (\bibinfo {year} {2016})}\BibitemShut
  {NoStop}%
\bibitem [{\citenamefont {Benenti}\ \emph {et~al.}(2017)\citenamefont
  {Benenti}, \citenamefont {Casati}, \citenamefont {Saito},\ and\ \citenamefont
  {Whitney}}]{BenentiPR17}%
  \BibitemOpen
  \bibfield  {author} {\bibinfo {author} {\bibfnamefont {G.}~\bibnamefont
  {Benenti}}, \bibinfo {author} {\bibfnamefont {G.}~\bibnamefont {Casati}},
  \bibinfo {author} {\bibfnamefont {K.}~\bibnamefont {Saito}}, \ and\ \bibinfo
  {author} {\bibfnamefont {R.~S.}\ \bibnamefont {Whitney}},\ }\bibfield
  {title} {\enquote {\bibinfo {title} {Fundamental aspects of steady-state
  conversion of heat to work at the nanoscale},}\ }\href {\doibase
  10.1016/j.physrep.2017.05.008} {\bibfield  {journal} {\bibinfo  {journal}
  {Phys. Rep.}\ }\textbf {\bibinfo {volume} {694}},\ \bibinfo {pages} {1 --
  124} (\bibinfo {year} {2017})}\BibitemShut {NoStop}%
\bibitem [{\citenamefont {Pekola}\ and\ \citenamefont
  {Karimi}(2021)}]{RMPPekola}%
  \BibitemOpen
  \bibfield  {author} {\bibinfo {author} {\bibfnamefont {J.~P.}\ \bibnamefont
  {Pekola}}\ and\ \bibinfo {author} {\bibfnamefont {B.}~\bibnamefont
  {Karimi}},\ }\bibfield  {title} {\enquote {\bibinfo {title} {Colloquium:
  Quantum heat transport in condensed matter systems},}\ }\href {\doibase
  10.1103/RevModPhys.93.041001} {\bibfield  {journal} {\bibinfo  {journal}
  {Rev. Mod. Phys.}\ }\textbf {\bibinfo {volume} {93}},\ \bibinfo {pages}
  {041001} (\bibinfo {year} {2021})}\BibitemShut {NoStop}%
\bibitem [{\citenamefont {Wang}\ \emph
  {et~al.}(2022{\natexlab{a}})\citenamefont {Wang}, \citenamefont {Wang},
  \citenamefont {Lu},\ and\ \citenamefont {Jiang}}]{MyReview}%
  \BibitemOpen
  \bibfield  {author} {\bibinfo {author} {\bibfnamefont {R.}~\bibnamefont
  {Wang}}, \bibinfo {author} {\bibfnamefont {C.}~\bibnamefont {Wang}}, \bibinfo
  {author} {\bibfnamefont {J.}~\bibnamefont {Lu}}, \ and\ \bibinfo {author}
  {\bibfnamefont {J.-H.}\ \bibnamefont {Jiang}},\ }\bibfield  {title} {\enquote
  {\bibinfo {title} {Inelastic thermoelectric transport and fluctuations in
  mesoscopic systems},}\ }\href {\doibase 10.1080/23746149.2022.2082317}
  {\bibfield  {journal} {\bibinfo  {journal} {Adv. Phys.: X}\ }\textbf
  {\bibinfo {volume} {7}},\ \bibinfo {pages} {2082317} (\bibinfo {year}
  {2022}{\natexlab{a}})}\BibitemShut {NoStop}%
\bibitem [{\citenamefont {Mukherjee}\ and\ \citenamefont
  {Divakaran}(2021)}]{Mukherjee21}%
  \BibitemOpen
  \bibfield  {author} {\bibinfo {author} {\bibfnamefont {V.}~\bibnamefont
  {Mukherjee}}\ and\ \bibinfo {author} {\bibfnamefont {U.}~\bibnamefont
  {Divakaran}},\ }\bibfield  {title} {\enquote {\bibinfo {title} {Many-body
  quantum thermal machines},}\ }\href {\doibase 10.1088/1361-648x/ac1b60}
  {\bibfield  {journal} {\bibinfo  {journal} {J. Phys.: Condens. Matter}\
  }\textbf {\bibinfo {volume} {33}},\ \bibinfo {pages} {454001} (\bibinfo
  {year} {2021})}\BibitemShut {NoStop}%
\bibitem [{\citenamefont {Whitney}(2014)}]{WhitneyPRL}%
  \BibitemOpen
  \bibfield  {author} {\bibinfo {author} {\bibfnamefont {R.~S.}\ \bibnamefont
  {Whitney}},\ }\bibfield  {title} {\enquote {\bibinfo {title} {Most efficient
  quantum thermoelectric at finite power output},}\ }\href {\doibase
  10.1103/PhysRevLett.112.130601} {\bibfield  {journal} {\bibinfo  {journal}
  {Phys. Rev. Lett.}\ }\textbf {\bibinfo {volume} {112}},\ \bibinfo {pages}
  {130601} (\bibinfo {year} {2014})}\BibitemShut {NoStop}%
\bibitem [{\citenamefont {Whitney}(2015)}]{WhitneyPRB}%
  \BibitemOpen
  \bibfield  {author} {\bibinfo {author} {\bibfnamefont {R.~S.}\ \bibnamefont
  {Whitney}},\ }\bibfield  {title} {\enquote {\bibinfo {title} {Finding the
  quantum thermoelectric with maximal efficiency and minimal entropy production
  at given power output},}\ }\href {\doibase 10.1103/PhysRevB.91.115425}
  {\bibfield  {journal} {\bibinfo  {journal} {Phys. Rev. B}\ }\textbf {\bibinfo
  {volume} {91}},\ \bibinfo {pages} {115425} (\bibinfo {year}
  {2015})}\BibitemShut {NoStop}%
\bibitem [{\citenamefont {Entin-Wohlman}\ \emph {et~al.}(2014)\citenamefont
  {Entin-Wohlman}, \citenamefont {Jiang},\ and\ \citenamefont
  {Imry}}]{JiangOra}%
  \BibitemOpen
  \bibfield  {author} {\bibinfo {author} {\bibfnamefont {O.}~\bibnamefont
  {Entin-Wohlman}}, \bibinfo {author} {\bibfnamefont {J.-H.}\ \bibnamefont
  {Jiang}}, \ and\ \bibinfo {author} {\bibfnamefont {Y.}~\bibnamefont {Imry}},\
  }\bibfield  {title} {\enquote {\bibinfo {title} {Efficiency and dissipation
  in a two-terminal thermoelectric junction, emphasizing small dissipation},}\
  }\href {\doibase 10.1103/PhysRevE.89.012123} {\bibfield  {journal} {\bibinfo
  {journal} {Phys. Rev. E}\ }\textbf {\bibinfo {volume} {89}},\ \bibinfo
  {pages} {012123} (\bibinfo {year} {2014})}\BibitemShut {NoStop}%
\bibitem [{\citenamefont {Josefsson}\ \emph {et~al.}(2019)\citenamefont
  {Josefsson}, \citenamefont {Svilans}, \citenamefont {Linke},\ and\
  \citenamefont {Leijnse}}]{Josefsson}%
  \BibitemOpen
  \bibfield  {author} {\bibinfo {author} {\bibfnamefont {M.}~\bibnamefont
  {Josefsson}}, \bibinfo {author} {\bibfnamefont {A.}~\bibnamefont {Svilans}},
  \bibinfo {author} {\bibfnamefont {H.}~\bibnamefont {Linke}}, \ and\ \bibinfo
  {author} {\bibfnamefont {M.}~\bibnamefont {Leijnse}},\ }\bibfield  {title}
  {\enquote {\bibinfo {title} {Optimal power and efficiency of single quantum
  dot heat engines: Theory and experiment},}\ }\href {\doibase
  10.1103/PhysRevB.99.235432} {\bibfield  {journal} {\bibinfo  {journal} {Phys.
  Rev. B}\ }\textbf {\bibinfo {volume} {99}},\ \bibinfo {pages} {235432}
  (\bibinfo {year} {2019})}\BibitemShut {NoStop}%
\bibitem [{\citenamefont {Zhai}\ \emph {et~al.}(2022)\citenamefont {Zhai},
  \citenamefont {Cui}, \citenamefont {Ma}, \citenamefont {Sun},\ and\
  \citenamefont {Dong}}]{zhai22}%
  \BibitemOpen
  \bibfield  {author} {\bibinfo {author} {\bibfnamefont {R.-X.}\ \bibnamefont
  {Zhai}}, \bibinfo {author} {\bibfnamefont {F.-M.}\ \bibnamefont {Cui}},
  \bibinfo {author} {\bibfnamefont {Y.-H.}\ \bibnamefont {Ma}}, \bibinfo
  {author} {\bibfnamefont {C.~P.}\ \bibnamefont {Sun}}, \ and\ \bibinfo
  {author} {\bibfnamefont {H.}~\bibnamefont {Dong}},\ }\bibfield  {title}
  {\enquote {\bibinfo {title} {Experimental implementation of finite-time
  carnot cycle},}\ }\href {https://arxiv.org/abs/2206.10153} {\bibfield
  {journal} {\bibinfo  {journal} {arXiv:2206.10153}\ } (\bibinfo {year}
  {2022})}\BibitemShut {NoStop}%
\bibitem [{\citenamefont {Proesmans}\ \emph
  {et~al.}(2016{\natexlab{a}})\citenamefont {Proesmans}, \citenamefont
  {Cleuren},\ and\ \citenamefont {Van~den Broeck}}]{Proesmans}%
  \BibitemOpen
  \bibfield  {author} {\bibinfo {author} {\bibfnamefont {K.}~\bibnamefont
  {Proesmans}}, \bibinfo {author} {\bibfnamefont {B.}~\bibnamefont {Cleuren}},
  \ and\ \bibinfo {author} {\bibfnamefont {C.}~\bibnamefont {Van~den Broeck}},\
  }\bibfield  {title} {\enquote {\bibinfo {title} {Power-efficiency-dissipation
  relations in linear thermodynamics},}\ }\href {\doibase
  10.1103/PhysRevLett.116.220601} {\bibfield  {journal} {\bibinfo  {journal}
  {Phys. Rev. Lett.}\ }\textbf {\bibinfo {volume} {116}},\ \bibinfo {pages}
  {220601} (\bibinfo {year} {2016}{\natexlab{a}})}\BibitemShut {NoStop}%
\bibitem [{\citenamefont {Shiraishi}\ \emph {et~al.}(2016)\citenamefont
  {Shiraishi}, \citenamefont {Saito},\ and\ \citenamefont {Tasaki}}]{Naoto}%
  \BibitemOpen
  \bibfield  {author} {\bibinfo {author} {\bibfnamefont {N.}~\bibnamefont
  {Shiraishi}}, \bibinfo {author} {\bibfnamefont {K.}~\bibnamefont {Saito}}, \
  and\ \bibinfo {author} {\bibfnamefont {H.}~\bibnamefont {Tasaki}},\
  }\bibfield  {title} {\enquote {\bibinfo {title} {Universal trade-off relation
  between power and efficiency for heat engines},}\ }\href {\doibase
  10.1103/PhysRevLett.117.190601} {\bibfield  {journal} {\bibinfo  {journal}
  {Phys. Rev. Lett.}\ }\textbf {\bibinfo {volume} {117}},\ \bibinfo {pages}
  {190601} (\bibinfo {year} {2016})}\BibitemShut {NoStop}%
\bibitem [{\citenamefont {Guo}\ \emph {et~al.}(2019)\citenamefont {Guo},
  \citenamefont {Liu},\ and\ \citenamefont {Yu}}]{YuPRE19}%
  \BibitemOpen
  \bibfield  {author} {\bibinfo {author} {\bibfnamefont {B.-q.}\ \bibnamefont
  {Guo}}, \bibinfo {author} {\bibfnamefont {T.}~\bibnamefont {Liu}}, \ and\
  \bibinfo {author} {\bibfnamefont {C.-s.}\ \bibnamefont {Yu}},\ }\bibfield
  {title} {\enquote {\bibinfo {title} {Multifunctional quantum thermal device
  utilizing three qubits},}\ }\href {\doibase 10.1103/PhysRevE.99.032112}
  {\bibfield  {journal} {\bibinfo  {journal} {Phys. Rev. E}\ }\textbf {\bibinfo
  {volume} {99}},\ \bibinfo {pages} {032112} (\bibinfo {year}
  {2019})}\BibitemShut {NoStop}%
\bibitem [{\citenamefont {Pietzonka}\ and\ \citenamefont
  {Seifert}(2018)}]{Constancy}%
  \BibitemOpen
  \bibfield  {author} {\bibinfo {author} {\bibfnamefont {P.}~\bibnamefont
  {Pietzonka}}\ and\ \bibinfo {author} {\bibfnamefont {U.}~\bibnamefont
  {Seifert}},\ }\bibfield  {title} {\enquote {\bibinfo {title} {Universal
  trade-off between power, efficiency, and constancy in steady-state heat
  engines},}\ }\href {\doibase 10.1103/PhysRevLett.120.190602} {\bibfield
  {journal} {\bibinfo  {journal} {Phys. Rev. Lett.}\ }\textbf {\bibinfo
  {volume} {120}},\ \bibinfo {pages} {190602} (\bibinfo {year}
  {2018})}\BibitemShut {NoStop}%
\bibitem [{\citenamefont {Ma}\ \emph {et~al.}(2020)\citenamefont {Ma},
  \citenamefont {Zhai}, \citenamefont {Chen}, \citenamefont {Sun},\ and\
  \citenamefont {Dong}}]{MaPRL20}%
  \BibitemOpen
  \bibfield  {author} {\bibinfo {author} {\bibfnamefont {Y.-H.}\ \bibnamefont
  {Ma}}, \bibinfo {author} {\bibfnamefont {R.-X.}\ \bibnamefont {Zhai}},
  \bibinfo {author} {\bibfnamefont {J.}~\bibnamefont {Chen}}, \bibinfo {author}
  {\bibfnamefont {C.~P.}\ \bibnamefont {Sun}}, \ and\ \bibinfo {author}
  {\bibfnamefont {H.}~\bibnamefont {Dong}},\ }\bibfield  {title} {\enquote
  {\bibinfo {title} {Experimental test of the $1/\ensuremath{\tau}$-scaling
  entropy generation in finite-time thermodynamics},}\ }\href {\doibase
  10.1103/PhysRevLett.125.210601} {\bibfield  {journal} {\bibinfo  {journal}
  {Phys. Rev. Lett.}\ }\textbf {\bibinfo {volume} {125}},\ \bibinfo {pages}
  {210601} (\bibinfo {year} {2020})}\BibitemShut {NoStop}%
\bibitem [{\citenamefont {Liu}\ \emph {et~al.}(2022)\citenamefont {Liu},
  \citenamefont {Yu},\ and\ \citenamefont {Yu}}]{YuEntropy}%
  \BibitemOpen
  \bibfield  {author} {\bibinfo {author} {\bibfnamefont {Y.-Q.}\ \bibnamefont
  {Liu}}, \bibinfo {author} {\bibfnamefont {D.-H.}\ \bibnamefont {Yu}}, \ and\
  \bibinfo {author} {\bibfnamefont {C.-S.}\ \bibnamefont {Yu}},\ }\bibfield
  {title} {\enquote {\bibinfo {title} {Common environmental effects on quantum
  thermal transistor},}\ }\href {\doibase 10.3390/e24010032} {\bibfield
  {journal} {\bibinfo  {journal} {Entropy}\ }\textbf {\bibinfo {volume} {24}}
  (\bibinfo {year} {2022}),\ 10.3390/e24010032}\BibitemShut {NoStop}%
\bibitem [{\citenamefont {Mahan}\ and\ \citenamefont {Sofo}(1996)}]{Mahan}%
  \BibitemOpen
  \bibfield  {author} {\bibinfo {author} {\bibfnamefont {G.~D.}\ \bibnamefont
  {Mahan}}\ and\ \bibinfo {author} {\bibfnamefont {J.~O.}\ \bibnamefont
  {Sofo}},\ }\bibfield  {title} {\enquote {\bibinfo {title} {The best
  thermoelectric},}\ }\href {\doibase 10.1073/pnas.93.15.7436} {\bibfield
  {journal} {\bibinfo  {journal} {Proc. Natl. Acad. Sci. USA}\ }\textbf
  {\bibinfo {volume} {93}},\ \bibinfo {pages} {7436--7439} (\bibinfo {year}
  {1996})}\BibitemShut {NoStop}%
\bibitem [{\citenamefont {Zhou}\ \emph {et~al.}(2011)\citenamefont {Zhou},
  \citenamefont {Yang}, \citenamefont {Chen},\ and\ \citenamefont
  {Dresselhaus}}]{ZhouPRL}%
  \BibitemOpen
  \bibfield  {author} {\bibinfo {author} {\bibfnamefont {J.}~\bibnamefont
  {Zhou}}, \bibinfo {author} {\bibfnamefont {R.}~\bibnamefont {Yang}}, \bibinfo
  {author} {\bibfnamefont {G.}~\bibnamefont {Chen}}, \ and\ \bibinfo {author}
  {\bibfnamefont {M.~S.}\ \bibnamefont {Dresselhaus}},\ }\bibfield  {title}
  {\enquote {\bibinfo {title} {Optimal bandwidth for high efficiency
  thermoelectrics},}\ }\href {\doibase 10.1103/PhysRevLett.107.226601}
  {\bibfield  {journal} {\bibinfo  {journal} {Phys. Rev. Lett.}\ }\textbf
  {\bibinfo {volume} {107}},\ \bibinfo {pages} {226601} (\bibinfo {year}
  {2011})}\BibitemShut {NoStop}%
\bibitem [{\citenamefont {Mazza}\ \emph {et~al.}(2015)\citenamefont {Mazza},
  \citenamefont {Valentini}, \citenamefont {Bosisio}, \citenamefont {Benenti},
  \citenamefont {Giovannetti}, \citenamefont {Fazio},\ and\ \citenamefont
  {Taddei}}]{MazzaPRB}%
  \BibitemOpen
  \bibfield  {author} {\bibinfo {author} {\bibfnamefont {F.}~\bibnamefont
  {Mazza}}, \bibinfo {author} {\bibfnamefont {S.}~\bibnamefont {Valentini}},
  \bibinfo {author} {\bibfnamefont {R.}~\bibnamefont {Bosisio}}, \bibinfo
  {author} {\bibfnamefont {G.}~\bibnamefont {Benenti}}, \bibinfo {author}
  {\bibfnamefont {V.}~\bibnamefont {Giovannetti}}, \bibinfo {author}
  {\bibfnamefont {R.}~\bibnamefont {Fazio}}, \ and\ \bibinfo {author}
  {\bibfnamefont {F.}~\bibnamefont {Taddei}},\ }\bibfield  {title} {\enquote
  {\bibinfo {title} {Separation of heat and charge currents for boosted
  thermoelectric conversion},}\ }\href {\doibase 10.1103/PhysRevB.91.245435}
  {\bibfield  {journal} {\bibinfo  {journal} {Phys. Rev. B}\ }\textbf {\bibinfo
  {volume} {91}},\ \bibinfo {pages} {245435} (\bibinfo {year}
  {2015})}\BibitemShut {NoStop}%
\bibitem [{\citenamefont {Entin-Wohlman}\ \emph {et~al.}(2010)\citenamefont
  {Entin-Wohlman}, \citenamefont {Imry},\ and\ \citenamefont
  {Aharony}}]{OraPRB2010}%
  \BibitemOpen
  \bibfield  {author} {\bibinfo {author} {\bibfnamefont {O.}~\bibnamefont
  {Entin-Wohlman}}, \bibinfo {author} {\bibfnamefont {Y.}~\bibnamefont {Imry}},
  \ and\ \bibinfo {author} {\bibfnamefont {A.}~\bibnamefont {Aharony}},\
  }\bibfield  {title} {\enquote {\bibinfo {title} {Three-terminal
  thermoelectric transport through a molecular junction},}\ }\href {\doibase
  10.1103/PhysRevB.82.115314} {\bibfield  {journal} {\bibinfo  {journal} {Phys.
  Rev. B}\ }\textbf {\bibinfo {volume} {82}},\ \bibinfo {pages} {115314}
  (\bibinfo {year} {2010})}\BibitemShut {NoStop}%
\bibitem [{\citenamefont {S\'anchez}\ and\ \citenamefont
  {L\'opez}(2013)}]{DavidPRL}%
  \BibitemOpen
  \bibfield  {author} {\bibinfo {author} {\bibfnamefont {D.}~\bibnamefont
  {S\'anchez}}\ and\ \bibinfo {author} {\bibfnamefont {R.}~\bibnamefont
  {L\'opez}},\ }\bibfield  {title} {\enquote {\bibinfo {title} {Scattering
  theory of nonlinear thermoelectric transport},}\ }\href {\doibase
  10.1103/PhysRevLett.110.026804} {\bibfield  {journal} {\bibinfo  {journal}
  {Phys. Rev. Lett.}\ }\textbf {\bibinfo {volume} {110}},\ \bibinfo {pages}
  {026804} (\bibinfo {year} {2013})}\BibitemShut {NoStop}%
\bibitem [{\citenamefont {Jiang}\ \emph
  {et~al.}(2013{\natexlab{a}})\citenamefont {Jiang}, \citenamefont
  {Entin-Wohlman},\ and\ \citenamefont {Imry}}]{JiangNJP}%
  \BibitemOpen
  \bibfield  {author} {\bibinfo {author} {\bibfnamefont {J.-H.}\ \bibnamefont
  {Jiang}}, \bibinfo {author} {\bibfnamefont {O.}~\bibnamefont
  {Entin-Wohlman}}, \ and\ \bibinfo {author} {\bibfnamefont {Y.}~\bibnamefont
  {Imry}},\ }\bibfield  {title} {\enquote {\bibinfo {title} {Three-terminal
  semiconductor junction thermoelectric devices: improving performance},}\
  }\href {http://stacks.iop.org/1367-2630/15/i=7/a=075021} {\bibfield
  {journal} {\bibinfo  {journal} {New J. Phys.}\ }\textbf {\bibinfo {volume}
  {15}},\ \bibinfo {pages} {075021} (\bibinfo {year}
  {2013}{\natexlab{a}})}\BibitemShut {NoStop}%
\bibitem [{\citenamefont {Sothmann}\ \emph {et~al.}(2013)\citenamefont
  {Sothmann}, \citenamefont {S{\'a}nchez}, \citenamefont {Jordan},\ and\
  \citenamefont {B{\"u}ttiker}}]{SothmannQW}%
  \BibitemOpen
  \bibfield  {author} {\bibinfo {author} {\bibfnamefont {B.}~\bibnamefont
  {Sothmann}}, \bibinfo {author} {\bibfnamefont {R.}~\bibnamefont
  {S{\'a}nchez}}, \bibinfo {author} {\bibfnamefont {A.~N}\ \bibnamefont
  {Jordan}}, \ and\ \bibinfo {author} {\bibfnamefont {M.}~\bibnamefont
  {B{\"u}ttiker}},\ }\bibfield  {title} {\enquote {\bibinfo {title} {Powerful
  energy harvester based on resonant-tunneling quantum wells},}\ }\href
  {http://stacks.iop.org/1367-2630/15/i=9/a=095021} {\bibfield  {journal}
  {\bibinfo  {journal} {New J. Phys.}\ }\textbf {\bibinfo {volume} {15}},\
  \bibinfo {pages} {095021} (\bibinfo {year} {2013})}\BibitemShut {NoStop}%
\bibitem [{\citenamefont {Yamamoto}\ \emph {et~al.}(2016)\citenamefont
  {Yamamoto}, \citenamefont {Entin-Wohlman}, \citenamefont {Aharony},\ and\
  \citenamefont {Hatano}}]{Yamamoto}%
  \BibitemOpen
  \bibfield  {author} {\bibinfo {author} {\bibfnamefont {K.}~\bibnamefont
  {Yamamoto}}, \bibinfo {author} {\bibfnamefont {O.}~\bibnamefont
  {Entin-Wohlman}}, \bibinfo {author} {\bibfnamefont {A.}~\bibnamefont
  {Aharony}}, \ and\ \bibinfo {author} {\bibfnamefont {N.}~\bibnamefont
  {Hatano}},\ }\bibfield  {title} {\enquote {\bibinfo {title} {Efficiency
  bounds on thermoelectric transport in magnetic fields: The role of inelastic
  processes},}\ }\href {\doibase 10.1103/PhysRevB.94.121402} {\bibfield
  {journal} {\bibinfo  {journal} {Phys. Rev. B}\ }\textbf {\bibinfo {volume}
  {94}},\ \bibinfo {pages} {121402} (\bibinfo {year} {2016})}\BibitemShut
  {NoStop}%
\bibitem [{\citenamefont {Zhang}\ \emph {et~al.}(2015)\citenamefont {Zhang},
  \citenamefont {Lin},\ and\ \citenamefont {Chen}}]{ChenPRE15}%
  \BibitemOpen
  \bibfield  {author} {\bibinfo {author} {\bibfnamefont {Y.}~\bibnamefont
  {Zhang}}, \bibinfo {author} {\bibfnamefont {G.}~\bibnamefont {Lin}}, \ and\
  \bibinfo {author} {\bibfnamefont {J.}~\bibnamefont {Chen}},\ }\bibfield
  {title} {\enquote {\bibinfo {title} {Three-terminal quantum-dot
  refrigerators},}\ }\href {\doibase 10.1103/PhysRevE.91.052118} {\bibfield
  {journal} {\bibinfo  {journal} {Phys. Rev. E}\ }\textbf {\bibinfo {volume}
  {91}},\ \bibinfo {pages} {052118} (\bibinfo {year} {2015})}\BibitemShut
  {NoStop}%
\bibitem [{\citenamefont {Agarwalla}\ \emph {et~al.}(2015)\citenamefont
  {Agarwalla}, \citenamefont {Jiang},\ and\ \citenamefont
  {Segal}}]{BijayJiang}%
  \BibitemOpen
  \bibfield  {author} {\bibinfo {author} {\bibfnamefont {B.~K.}\ \bibnamefont
  {Agarwalla}}, \bibinfo {author} {\bibfnamefont {J.-H.}\ \bibnamefont
  {Jiang}}, \ and\ \bibinfo {author} {\bibfnamefont {D.}~\bibnamefont
  {Segal}},\ }\bibfield  {title} {\enquote {\bibinfo {title} {Full counting
  statistics of vibrationally assisted electronic conduction: Transport and
  fluctuations of thermoelectric efficiency},}\ }\href {\doibase
  10.1103/PhysRevB.92.245418} {\bibfield  {journal} {\bibinfo  {journal} {Phys.
  Rev. B}\ }\textbf {\bibinfo {volume} {92}},\ \bibinfo {pages} {245418}
  (\bibinfo {year} {2015})}\BibitemShut {NoStop}%
\bibitem [{\citenamefont {Agarwalla}\ \emph {et~al.}(2017)\citenamefont
  {Agarwalla}, \citenamefont {Jiang},\ and\ \citenamefont
  {Segal}}]{JiangBijayPRB17}%
  \BibitemOpen
  \bibfield  {author} {\bibinfo {author} {\bibfnamefont {B.~K.}\ \bibnamefont
  {Agarwalla}}, \bibinfo {author} {\bibfnamefont {J.-H.}\ \bibnamefont
  {Jiang}}, \ and\ \bibinfo {author} {\bibfnamefont {D.}~\bibnamefont
  {Segal}},\ }\bibfield  {title} {\enquote {\bibinfo {title} {Quantum
  efficiency bound for continuous heat engines coupled to noncanonical
  reservoirs},}\ }\href {\doibase 10.1103/PhysRevB.96.104304} {\bibfield
  {journal} {\bibinfo  {journal} {Phys. Rev. B}\ }\textbf {\bibinfo {volume}
  {96}},\ \bibinfo {pages} {104304} (\bibinfo {year} {2017})}\BibitemShut
  {NoStop}%
\bibitem [{\citenamefont {Wang}\ \emph
  {et~al.}(2022{\natexlab{b}})\citenamefont {Wang}, \citenamefont {Wang},
  \citenamefont {Wang},\ and\ \citenamefont {Ren}}]{WangPRL}%
  \BibitemOpen
  \bibfield  {author} {\bibinfo {author} {\bibfnamefont {L.}~\bibnamefont
  {Wang}}, \bibinfo {author} {\bibfnamefont {Z.}~\bibnamefont {Wang}}, \bibinfo
  {author} {\bibfnamefont {C.}~\bibnamefont {Wang}}, \ and\ \bibinfo {author}
  {\bibfnamefont {J.}~\bibnamefont {Ren}},\ }\bibfield  {title} {\enquote
  {\bibinfo {title} {Cycle flux ranking of network analysis in quantum thermal
  devices},}\ }\href {\doibase 10.1103/PhysRevLett.128.067701} {\bibfield
  {journal} {\bibinfo  {journal} {Phys. Rev. Lett.}\ }\textbf {\bibinfo
  {volume} {128}},\ \bibinfo {pages} {067701} (\bibinfo {year}
  {2022}{\natexlab{b}})}\BibitemShut {NoStop}%
\bibitem [{\citenamefont {S\'anchez}\ and\ \citenamefont
  {B\"uttiker}(2011)}]{Rafael}%
  \BibitemOpen
  \bibfield  {author} {\bibinfo {author} {\bibfnamefont {R.}~\bibnamefont
  {S\'anchez}}\ and\ \bibinfo {author} {\bibfnamefont {M.}~\bibnamefont
  {B\"uttiker}},\ }\bibfield  {title} {\enquote {\bibinfo {title} {Optimal
  energy quanta to current conversion},}\ }\href {\doibase
  10.1103/PhysRevB.83.085428} {\bibfield  {journal} {\bibinfo  {journal} {Phys.
  Rev. B}\ }\textbf {\bibinfo {volume} {83}},\ \bibinfo {pages} {085428}
  (\bibinfo {year} {2011})}\BibitemShut {NoStop}%
\bibitem [{\citenamefont {Thierschmann}\ \emph {et~al.}(2015)\citenamefont
  {Thierschmann}, \citenamefont {Arnold}, \citenamefont {Mittermüller},
  \citenamefont {Maier}, \citenamefont {Heyn}, \citenamefont {Hansen},
  \citenamefont {Buhmann},\ and\ \citenamefont {Molenkamp}}]{ThiersNJP2015}%
  \BibitemOpen
  \bibfield  {author} {\bibinfo {author} {\bibfnamefont {H.}~\bibnamefont
  {Thierschmann}}, \bibinfo {author} {\bibfnamefont {F.}~\bibnamefont
  {Arnold}}, \bibinfo {author} {\bibfnamefont {M.}~\bibnamefont
  {Mittermüller}}, \bibinfo {author} {\bibfnamefont {L.}~\bibnamefont
  {Maier}}, \bibinfo {author} {\bibfnamefont {C.}~\bibnamefont {Heyn}},
  \bibinfo {author} {\bibfnamefont {W.}~\bibnamefont {Hansen}}, \bibinfo
  {author} {\bibfnamefont {H.}~\bibnamefont {Buhmann}}, \ and\ \bibinfo
  {author} {\bibfnamefont {L.~W.}\ \bibnamefont {Molenkamp}},\ }\bibfield
  {title} {\enquote {\bibinfo {title} {Thermal gating of charge currents with
  coulomb coupled quantum dots},}\ }\href {\doibase
  10.1088/1367-2630/17/11/113003} {\bibfield  {journal} {\bibinfo  {journal}
  {New J. Phys.}\ }\textbf {\bibinfo {volume} {17}},\ \bibinfo {pages} {113003}
  (\bibinfo {year} {2015})}\BibitemShut {NoStop}%
\bibitem [{\citenamefont {Tabatabaei}\ \emph {et~al.}(2020)\citenamefont
  {Tabatabaei}, \citenamefont {S\'anchez}, \citenamefont {Yeyati},\ and\
  \citenamefont {S\'anchez}}]{TabatabaeiPRL20}%
  \BibitemOpen
  \bibfield  {author} {\bibinfo {author} {\bibfnamefont {S.~M.}\ \bibnamefont
  {Tabatabaei}}, \bibinfo {author} {\bibfnamefont {D.}~\bibnamefont
  {S\'anchez}}, \bibinfo {author} {\bibfnamefont {Alfredo~L.}\ \bibnamefont
  {Yeyati}}, \ and\ \bibinfo {author} {\bibfnamefont {R.}~\bibnamefont
  {S\'anchez}},\ }\bibfield  {title} {\enquote {\bibinfo {title}
  {Andreev-coulomb drag in coupled quantum dots},}\ }\href {\doibase
  10.1103/PhysRevLett.125.247701} {\bibfield  {journal} {\bibinfo  {journal}
  {Phys. Rev. Lett.}\ }\textbf {\bibinfo {volume} {125}},\ \bibinfo {pages}
  {247701} (\bibinfo {year} {2020})}\BibitemShut {NoStop}%
\bibitem [{\citenamefont {Jiang}\ \emph {et~al.}(2012)\citenamefont {Jiang},
  \citenamefont {Entin-Wohlman},\ and\ \citenamefont {Imry}}]{Jiang2012}%
  \BibitemOpen
  \bibfield  {author} {\bibinfo {author} {\bibfnamefont {J.-H.}\ \bibnamefont
  {Jiang}}, \bibinfo {author} {\bibfnamefont {O.}~\bibnamefont
  {Entin-Wohlman}}, \ and\ \bibinfo {author} {\bibfnamefont {Y.}~\bibnamefont
  {Imry}},\ }\bibfield  {title} {\enquote {\bibinfo {title} {Thermoelectric
  three-terminal hopping transport through one-dimensional nanosystems},}\
  }\href {\doibase 10.1103/PhysRevB.85.075412} {\bibfield  {journal} {\bibinfo
  {journal} {Phys. Rev. B}\ }\textbf {\bibinfo {volume} {85}},\ \bibinfo
  {pages} {075412} (\bibinfo {year} {2012})}\BibitemShut {NoStop}%
\bibitem [{\citenamefont {Jiang}\ \emph
  {et~al.}(2013{\natexlab{b}})\citenamefont {Jiang}, \citenamefont
  {Entin-Wohlman},\ and\ \citenamefont {Imry}}]{Jiang2013}%
  \BibitemOpen
  \bibfield  {author} {\bibinfo {author} {\bibfnamefont {J.-H.}\ \bibnamefont
  {Jiang}}, \bibinfo {author} {\bibfnamefont {O.}~\bibnamefont
  {Entin-Wohlman}}, \ and\ \bibinfo {author} {\bibfnamefont {Y.}~\bibnamefont
  {Imry}},\ }\bibfield  {title} {\enquote {\bibinfo {title} {Hopping
  thermoelectric transport in finite systems: Boundary effects},}\ }\href
  {\doibase 10.1103/PhysRevB.87.205420} {\bibfield  {journal} {\bibinfo
  {journal} {Phys. Rev. B}\ }\textbf {\bibinfo {volume} {87}},\ \bibinfo
  {pages} {205420} (\bibinfo {year} {2013}{\natexlab{b}})}\BibitemShut
  {NoStop}%
\bibitem [{\citenamefont {Cleuren}\ \emph {et~al.}(2012)\citenamefont
  {Cleuren}, \citenamefont {Rutten},\ and\ \citenamefont {Van~den
  Broeck}}]{Cooling2}%
  \BibitemOpen
  \bibfield  {author} {\bibinfo {author} {\bibfnamefont {B.}~\bibnamefont
  {Cleuren}}, \bibinfo {author} {\bibfnamefont {B.}~\bibnamefont {Rutten}}, \
  and\ \bibinfo {author} {\bibfnamefont {C.}~\bibnamefont {Van~den Broeck}},\
  }\bibfield  {title} {\enquote {\bibinfo {title} {Cooling by heating:
  Refrigeration powered by photons},}\ }\href {\doibase
  10.1103/PhysRevLett.108.120603} {\bibfield  {journal} {\bibinfo  {journal}
  {Phys. Rev. Lett.}\ }\textbf {\bibinfo {volume} {108}},\ \bibinfo {pages}
  {120603} (\bibinfo {year} {2012})}\BibitemShut {NoStop}%
\bibitem [{\citenamefont {Lu}\ \emph {et~al.}(2019{\natexlab{a}})\citenamefont
  {Lu}, \citenamefont {Wang}, \citenamefont {Ren}, \citenamefont {Kulkarni},\
  and\ \citenamefont {Jiang}}]{MyPRBdiode}%
  \BibitemOpen
  \bibfield  {author} {\bibinfo {author} {\bibfnamefont {J.}~\bibnamefont
  {Lu}}, \bibinfo {author} {\bibfnamefont {R.}~\bibnamefont {Wang}}, \bibinfo
  {author} {\bibfnamefont {J.}~\bibnamefont {Ren}}, \bibinfo {author}
  {\bibfnamefont {M.}~\bibnamefont {Kulkarni}}, \ and\ \bibinfo {author}
  {\bibfnamefont {J.-H.}\ \bibnamefont {Jiang}},\ }\bibfield  {title} {\enquote
  {\bibinfo {title} {Quantum-dot circuit-qed thermoelectric diodes and
  transistors},}\ }\href {\doibase 10.1103/PhysRevB.99.035129} {\bibfield
  {journal} {\bibinfo  {journal} {Phys. Rev. B}\ }\textbf {\bibinfo {volume}
  {99}},\ \bibinfo {pages} {035129} (\bibinfo {year}
  {2019}{\natexlab{a}})}\BibitemShut {NoStop}%
\bibitem [{\citenamefont {Kosloff}\ and\ \citenamefont
  {Levy}(2014)}]{ReviewKosloff}%
  \BibitemOpen
  \bibfield  {author} {\bibinfo {author} {\bibfnamefont {R.}~\bibnamefont
  {Kosloff}}\ and\ \bibinfo {author} {\bibfnamefont {A.}~\bibnamefont {Levy}},\
  }\bibfield  {title} {\enquote {\bibinfo {title} {Quantum heat engines and
  refrigerators: Continuous devices},}\ }\href {\doibase
  10.1146/annurev-physchem-040513-103724} {\bibfield  {journal} {\bibinfo
  {journal} {Annu. Rev. Phys. Chem.}\ }\textbf {\bibinfo {volume} {65}},\
  \bibinfo {pages} {365--393} (\bibinfo {year} {2014})}\BibitemShut {NoStop}%
\bibitem [{\citenamefont {Wang}\ \emph {et~al.}(2018)\citenamefont {Wang},
  \citenamefont {Lu}, \citenamefont {Wang},\ and\ \citenamefont
  {Jiang}}]{Rongqian}%
  \BibitemOpen
  \bibfield  {author} {\bibinfo {author} {\bibfnamefont {R.}~\bibnamefont
  {Wang}}, \bibinfo {author} {\bibfnamefont {J.}~\bibnamefont {Lu}}, \bibinfo
  {author} {\bibfnamefont {C.}~\bibnamefont {Wang}}, \ and\ \bibinfo {author}
  {\bibfnamefont {J.-H.}\ \bibnamefont {Jiang}},\ }\bibfield  {title} {\enquote
  {\bibinfo {title} {Nonlinear effects for three-terminal heat engine and
  refrigerator},}\ }\href {\doibase 10.1038/s41598-018-20757-8} {\bibfield
  {journal} {\bibinfo  {journal} {Sci. Rep.}\ }\textbf {\bibinfo {volume}
  {8}},\ \bibinfo {pages} {2607} (\bibinfo {year} {2018})}\BibitemShut
  {NoStop}%
\bibitem [{\citenamefont {S\'anchez}\ \emph {et~al.}(2019)\citenamefont
  {S\'anchez}, \citenamefont {S\'anchez}, \citenamefont {L\'opez},\ and\
  \citenamefont {Sothmann}}]{David-refrigerator}%
  \BibitemOpen
  \bibfield  {author} {\bibinfo {author} {\bibfnamefont {D.}~\bibnamefont
  {S\'anchez}}, \bibinfo {author} {\bibfnamefont {R.}~\bibnamefont
  {S\'anchez}}, \bibinfo {author} {\bibfnamefont {R.}~\bibnamefont {L\'opez}},
  \ and\ \bibinfo {author} {\bibfnamefont {B.}~\bibnamefont {Sothmann}},\
  }\bibfield  {title} {\enquote {\bibinfo {title} {Nonlinear chiral
  refrigerators},}\ }\href {\doibase 10.1103/PhysRevB.99.245304} {\bibfield
  {journal} {\bibinfo  {journal} {Phys. Rev. B}\ }\textbf {\bibinfo {volume}
  {99}},\ \bibinfo {pages} {245304} (\bibinfo {year} {2019})}\BibitemShut
  {NoStop}%
\bibitem [{\citenamefont {Ren}\ \emph {et~al.}(2010)\citenamefont {Ren},
  \citenamefont {H\"anggi},\ and\ \citenamefont {Li}}]{RenPRL10}%
  \BibitemOpen
  \bibfield  {author} {\bibinfo {author} {\bibfnamefont {J.}~\bibnamefont
  {Ren}}, \bibinfo {author} {\bibfnamefont {P.}~\bibnamefont {H\"anggi}}, \
  and\ \bibinfo {author} {\bibfnamefont {B.}~\bibnamefont {Li}},\ }\bibfield
  {title} {\enquote {\bibinfo {title} {Berry-phase-induced heat pumping and its
  impact on the fluctuation theorem},}\ }\href {\doibase
  10.1103/PhysRevLett.104.170601} {\bibfield  {journal} {\bibinfo  {journal}
  {Phys. Rev. Lett.}\ }\textbf {\bibinfo {volume} {104}},\ \bibinfo {pages}
  {170601} (\bibinfo {year} {2010})}\BibitemShut {NoStop}%
\bibitem [{\citenamefont {Wang}\ \emph
  {et~al.}(2022{\natexlab{c}})\citenamefont {Wang}, \citenamefont {Wang},
  \citenamefont {Chen}, \citenamefont {Wang},\ and\ \citenamefont
  {Ren}}]{wangpump}%
  \BibitemOpen
  \bibfield  {author} {\bibinfo {author} {\bibfnamefont {Z.}~\bibnamefont
  {Wang}}, \bibinfo {author} {\bibfnamefont {L.}~\bibnamefont {Wang}}, \bibinfo
  {author} {\bibfnamefont {J.}~\bibnamefont {Chen}}, \bibinfo {author}
  {\bibfnamefont {C.}~\bibnamefont {Wang}}, \ and\ \bibinfo {author}
  {\bibfnamefont {J.}~\bibnamefont {Ren}},\ }\bibfield  {title} {\enquote
  {\bibinfo {title} {Geometric heat pump: Controlling thermal transport with
  time-dependent modulations},}\ }\href {\doibase 10.1007/s11467-021-1095-4}
  {\bibfield  {journal} {\bibinfo  {journal} {Front. Phys.}\ }\textbf {\bibinfo
  {volume} {17}},\ \bibinfo {pages} {1--14} (\bibinfo {year}
  {2022}{\natexlab{c}})}\BibitemShut {NoStop}%
\bibitem [{\citenamefont {Wang}\ \emph
  {et~al.}(2022{\natexlab{d}})\citenamefont {Wang}, \citenamefont {Chen},\ and\
  \citenamefont {Ren}}]{WangZiPRE}%
  \BibitemOpen
  \bibfield  {author} {\bibinfo {author} {\bibfnamefont {Z.}~\bibnamefont
  {Wang}}, \bibinfo {author} {\bibfnamefont {J.}~\bibnamefont {Chen}}, \ and\
  \bibinfo {author} {\bibfnamefont {J.}~\bibnamefont {Ren}},\ }\bibfield
  {title} {\enquote {\bibinfo {title} {Geometric heat pump and no-go
  restrictions of nonreciprocity in modulated thermal diffusion},}\ }\href
  {\doibase 10.1103/PhysRevE.106.L032102} {\bibfield  {journal} {\bibinfo
  {journal} {Phys. Rev. E}\ }\textbf {\bibinfo {volume} {106}},\ \bibinfo
  {pages} {L032102} (\bibinfo {year} {2022}{\natexlab{d}})}\BibitemShut
  {NoStop}%
\bibitem [{\citenamefont {Myers}\ \emph {et~al.}(2022)\citenamefont {Myers},
  \citenamefont {Abah},\ and\ \citenamefont {Deffner}}]{AVS}%
  \BibitemOpen
  \bibfield  {author} {\bibinfo {author} {\bibfnamefont {N.~M.}\ \bibnamefont
  {Myers}}, \bibinfo {author} {\bibfnamefont {O.}~\bibnamefont {Abah}}, \ and\
  \bibinfo {author} {\bibfnamefont {S.}~\bibnamefont {Deffner}},\ }\bibfield
  {title} {\enquote {\bibinfo {title} {Quantum thermodynamic devices: From
  theoretical proposals to experimental reality},}\ }\href {\doibase
  10.1116/5.0083192} {\bibfield  {journal} {\bibinfo  {journal} {AVS Quantum
  Sci.}\ }\textbf {\bibinfo {volume} {4}},\ \bibinfo {pages} {027101} (\bibinfo
  {year} {2022})}\BibitemShut {NoStop}%
\bibitem [{\citenamefont {Manzano}\ and\ \citenamefont
  {Zambrini}(2022)}]{ManzanoReview}%
  \BibitemOpen
  \bibfield  {author} {\bibinfo {author} {\bibfnamefont {G.}~\bibnamefont
  {Manzano}}\ and\ \bibinfo {author} {\bibfnamefont {R.}~\bibnamefont
  {Zambrini}},\ }\bibfield  {title} {\enquote {\bibinfo {title} {Quantum
  thermodynamics under continuous monitoring: A general framework},}\ }\href
  {\doibase 10.1116/5.0079886} {\bibfield  {journal} {\bibinfo  {journal} {AVS
  Quantum Science}\ }\textbf {\bibinfo {volume} {4}},\ \bibinfo {pages}
  {025302} (\bibinfo {year} {2022})}\BibitemShut {NoStop}%
\bibitem [{\citenamefont {Tu}(2021)}]{tu21}%
  \BibitemOpen
  \bibfield  {author} {\bibinfo {author} {\bibfnamefont {Z.-C.}\ \bibnamefont
  {Tu}},\ }\bibfield  {title} {\enquote {\bibinfo {title} {Abstract models for
  heat engines},}\ }\href {\doibase 10.1007/s11467-020-1029-6} {\bibfield
  {journal} {\bibinfo  {journal} {Front. Phys.}\ }\textbf {\bibinfo {volume}
  {16}},\ \bibinfo {pages} {1--12} (\bibinfo {year} {2021})}\BibitemShut
  {NoStop}%
\bibitem [{\citenamefont {Jiang}\ and\ \citenamefont {Imry}(2017)}]{Jiang2017}%
  \BibitemOpen
  \bibfield  {author} {\bibinfo {author} {\bibfnamefont {J.-H.}\ \bibnamefont
  {Jiang}}\ and\ \bibinfo {author} {\bibfnamefont {Y.}~\bibnamefont {Imry}},\
  }\bibfield  {title} {\enquote {\bibinfo {title} {Enhancing thermoelectric
  performance using nonlinear transport effects},}\ }\href {\doibase
  10.1103/PhysRevApplied.7.064001} {\bibfield  {journal} {\bibinfo  {journal}
  {Phys. Rev. Applied}\ }\textbf {\bibinfo {volume} {7}},\ \bibinfo {pages}
  {064001} (\bibinfo {year} {2017})}\BibitemShut {NoStop}%
\bibitem [{\citenamefont {Jiang}\ and\ \citenamefont
  {Imry}(2018)}]{JiangNearfield}%
  \BibitemOpen
  \bibfield  {author} {\bibinfo {author} {\bibfnamefont {J.-H.}\ \bibnamefont
  {Jiang}}\ and\ \bibinfo {author} {\bibfnamefont {Y.}~\bibnamefont {Imry}},\
  }\bibfield  {title} {\enquote {\bibinfo {title} {Near-field three-terminal
  thermoelectric heat engine},}\ }\href {\doibase 10.1103/PhysRevB.97.125422}
  {\bibfield  {journal} {\bibinfo  {journal} {Phys. Rev. B}\ }\textbf {\bibinfo
  {volume} {97}},\ \bibinfo {pages} {125422} (\bibinfo {year}
  {2018})}\BibitemShut {NoStop}%
\bibitem [{\citenamefont {Lu}\ \emph {et~al.}(2022)\citenamefont {Lu},
  \citenamefont {Wang}, \citenamefont {Peng}, \citenamefont {Wang},
  \citenamefont {Jiang},\ and\ \citenamefont {Ren}}]{MyPRBGTUR}%
  \BibitemOpen
  \bibfield  {author} {\bibinfo {author} {\bibfnamefont {J.}~\bibnamefont
  {Lu}}, \bibinfo {author} {\bibfnamefont {Z.}~\bibnamefont {Wang}}, \bibinfo
  {author} {\bibfnamefont {J.}~\bibnamefont {Peng}}, \bibinfo {author}
  {\bibfnamefont {C.}~\bibnamefont {Wang}}, \bibinfo {author} {\bibfnamefont
  {J.-H.}\ \bibnamefont {Jiang}}, \ and\ \bibinfo {author} {\bibfnamefont
  {J.}~\bibnamefont {Ren}},\ }\bibfield  {title} {\enquote {\bibinfo {title}
  {Geometric thermodynamic uncertainty relation in a periodically driven
  thermoelectric heat engine},}\ }\href {\doibase 10.1103/PhysRevB.105.115428}
  {\bibfield  {journal} {\bibinfo  {journal} {Phys. Rev. B}\ }\textbf {\bibinfo
  {volume} {105}},\ \bibinfo {pages} {115428} (\bibinfo {year}
  {2022})}\BibitemShut {NoStop}%
\bibitem [{\citenamefont {Entin-Wohlman}\ \emph {et~al.}(2015)\citenamefont
  {Entin-Wohlman}, \citenamefont {Imry},\ and\ \citenamefont
  {Aharony}}]{Ora2015}%
  \BibitemOpen
  \bibfield  {author} {\bibinfo {author} {\bibfnamefont {O.}~\bibnamefont
  {Entin-Wohlman}}, \bibinfo {author} {\bibfnamefont {Y.}~\bibnamefont {Imry}},
  \ and\ \bibinfo {author} {\bibfnamefont {A.}~\bibnamefont {Aharony}},\
  }\bibfield  {title} {\enquote {\bibinfo {title} {Enhanced performance of
  joint cooling and energy production},}\ }\href {\doibase
  10.1103/PhysRevB.91.054302} {\bibfield  {journal} {\bibinfo  {journal} {Phys.
  Rev. B}\ }\textbf {\bibinfo {volume} {91}},\ \bibinfo {pages} {054302}
  (\bibinfo {year} {2015})}\BibitemShut {NoStop}%
\bibitem [{\citenamefont {Manzano}\ \emph {et~al.}(2020)\citenamefont
  {Manzano}, \citenamefont {S\'anchez}, \citenamefont {Silva}, \citenamefont
  {Haack}, \citenamefont {Brask}, \citenamefont {Brunner},\ and\ \citenamefont
  {Potts}}]{ManzanoPRR}%
  \BibitemOpen
  \bibfield  {author} {\bibinfo {author} {\bibfnamefont {G.}~\bibnamefont
  {Manzano}}, \bibinfo {author} {\bibfnamefont {R.}~\bibnamefont {S\'anchez}},
  \bibinfo {author} {\bibfnamefont {R.}~\bibnamefont {Silva}}, \bibinfo
  {author} {\bibfnamefont {G.}~\bibnamefont {Haack}}, \bibinfo {author}
  {\bibfnamefont {J.~B.}\ \bibnamefont {Brask}}, \bibinfo {author}
  {\bibfnamefont {N.}~\bibnamefont {Brunner}}, \ and\ \bibinfo {author}
  {\bibfnamefont {P.~P.}\ \bibnamefont {Potts}},\ }\bibfield  {title} {\enquote
  {\bibinfo {title} {Hybrid thermal machines: Generalized thermodynamic
  resources for multitasking},}\ }\href {\doibase
  10.1103/PhysRevResearch.2.043302} {\bibfield  {journal} {\bibinfo  {journal}
  {Phys. Rev. Research}\ }\textbf {\bibinfo {volume} {2}},\ \bibinfo {pages}
  {043302} (\bibinfo {year} {2020})}\BibitemShut {NoStop}%
\bibitem [{\citenamefont {Hammam}\ \emph {et~al.}(2022)\citenamefont {Hammam},
  \citenamefont {Leitch}, \citenamefont {Hassouni},\ and\ \citenamefont
  {De~Chiara}}]{hammam22}%
  \BibitemOpen
  \bibfield  {author} {\bibinfo {author} {\bibfnamefont {K.}~\bibnamefont
  {Hammam}}, \bibinfo {author} {\bibfnamefont {H.}~\bibnamefont {Leitch}},
  \bibinfo {author} {\bibfnamefont {Y.}~\bibnamefont {Hassouni}}, \ and\
  \bibinfo {author} {\bibfnamefont {G.}~\bibnamefont {De~Chiara}},\ }\bibfield
  {title} {\enquote {\bibinfo {title} {Exploiting coherence for quantum
  thermodynamic advantage},}\ }\href {https://arxiv.org/abs/2202.07515}
  {\bibfield  {journal} {\bibinfo  {journal} {arXiv:2202.07515}\ } (\bibinfo
  {year} {2022})}\BibitemShut {NoStop}%
\bibitem [{\citenamefont {Landi}\ and\ \citenamefont
  {Paternostro}(2021)}]{RMPLandi}%
  \BibitemOpen
  \bibfield  {author} {\bibinfo {author} {\bibfnamefont {G.~T.}\ \bibnamefont
  {Landi}}\ and\ \bibinfo {author} {\bibfnamefont {M.}~\bibnamefont
  {Paternostro}},\ }\bibfield  {title} {\enquote {\bibinfo {title}
  {Irreversible entropy production: From classical to quantum},}\ }\href
  {\doibase 10.1103/RevModPhys.93.035008} {\bibfield  {journal} {\bibinfo
  {journal} {Rev. Mod. Phys.}\ }\textbf {\bibinfo {volume} {93}},\ \bibinfo
  {pages} {035008} (\bibinfo {year} {2021})}\BibitemShut {NoStop}%
\bibitem [{\citenamefont {Marcella}(1992)}]{Thomas92}%
  \BibitemOpen
  \bibfield  {author} {\bibinfo {author} {\bibfnamefont {Thomas~V.}\
  \bibnamefont {Marcella}},\ }\bibfield  {title} {\enquote {\bibinfo {title}
  {Entropy production and the second law of thermodynamics: An introduction to
  second law analysis},}\ }\href {\doibase 10.1119/1.17138} {\bibfield
  {journal} {\bibinfo  {journal} {Am. J. Phys.}\ }\textbf {\bibinfo {volume}
  {60}},\ \bibinfo {pages} {888--895} (\bibinfo {year} {1992})}\BibitemShut
  {NoStop}%
\bibitem [{\citenamefont {Mukherjee}\ \emph {et~al.}(2020)\citenamefont
  {Mukherjee}, \citenamefont {De},\ and\ \citenamefont
  {Muralidharan}}]{MukherjeeJAP}%
  \BibitemOpen
  \bibfield  {author} {\bibinfo {author} {\bibfnamefont {S.}~\bibnamefont
  {Mukherjee}}, \bibinfo {author} {\bibfnamefont {B.}~\bibnamefont {De}}, \
  and\ \bibinfo {author} {\bibfnamefont {B.}~\bibnamefont {Muralidharan}},\
  }\bibfield  {title} {\enquote {\bibinfo {title} {Three-terminal
  vibron-coupled hybrid quantum dot thermoelectric refrigeration},}\ }\href
  {\doibase 10.1063/5.0032215} {\bibfield  {journal} {\bibinfo  {journal} {J.
  Appl. Phys.}\ }\textbf {\bibinfo {volume} {128}},\ \bibinfo {pages} {234303}
  (\bibinfo {year} {2020})}\BibitemShut {NoStop}%
\bibitem [{\citenamefont {Hajiloo}\ \emph {et~al.}(2020)\citenamefont
  {Hajiloo}, \citenamefont {S\'anchez}, \citenamefont {Whitney},\ and\
  \citenamefont {Splettstoesser}}]{HajilooPRB}%
  \BibitemOpen
  \bibfield  {author} {\bibinfo {author} {\bibfnamefont {F.}~\bibnamefont
  {Hajiloo}}, \bibinfo {author} {\bibfnamefont {R.}~\bibnamefont {S\'anchez}},
  \bibinfo {author} {\bibfnamefont {R.~S.}\ \bibnamefont {Whitney}}, \ and\
  \bibinfo {author} {\bibfnamefont {J.}~\bibnamefont {Splettstoesser}},\
  }\bibfield  {title} {\enquote {\bibinfo {title} {Quantifying nonequilibrium
  thermodynamic operations in a multiterminal mesoscopic system},}\ }\href
  {\doibase 10.1103/PhysRevB.102.155405} {\bibfield  {journal} {\bibinfo
  {journal} {Phys. Rev. B}\ }\textbf {\bibinfo {volume} {102}},\ \bibinfo
  {pages} {155405} (\bibinfo {year} {2020})}\BibitemShut {NoStop}%
\bibitem [{\citenamefont {Carrega}\ \emph {et~al.}(2022)\citenamefont
  {Carrega}, \citenamefont {Cangemi}, \citenamefont {De~Filippis},
  \citenamefont {Cataudella}, \citenamefont {Benenti},\ and\ \citenamefont
  {Sassetti}}]{CarregaPRXQ}%
  \BibitemOpen
  \bibfield  {author} {\bibinfo {author} {\bibfnamefont {M.}~\bibnamefont
  {Carrega}}, \bibinfo {author} {\bibfnamefont {L.~M.}\ \bibnamefont
  {Cangemi}}, \bibinfo {author} {\bibfnamefont {G.}~\bibnamefont
  {De~Filippis}}, \bibinfo {author} {\bibfnamefont {V.}~\bibnamefont
  {Cataudella}}, \bibinfo {author} {\bibfnamefont {G.}~\bibnamefont {Benenti}},
  \ and\ \bibinfo {author} {\bibfnamefont {M.}~\bibnamefont {Sassetti}},\
  }\bibfield  {title} {\enquote {\bibinfo {title} {Engineering dynamical
  couplings for quantum thermodynamic tasks},}\ }\href {\doibase
  10.1103/PRXQuantum.3.010323} {\bibfield  {journal} {\bibinfo  {journal} {PRX
  Quantum}\ }\textbf {\bibinfo {volume} {3}},\ \bibinfo {pages} {010323}
  (\bibinfo {year} {2022})}\BibitemShut {NoStop}%
\bibitem [{\citenamefont {Strasberg}\ and\ \citenamefont
  {Winter}(2021{\natexlab{a}})}]{PRXQ}%
  \BibitemOpen
  \bibfield  {author} {\bibinfo {author} {\bibfnamefont {P.}~\bibnamefont
  {Strasberg}}\ and\ \bibinfo {author} {\bibfnamefont {A.}~\bibnamefont
  {Winter}},\ }\bibfield  {title} {\enquote {\bibinfo {title} {First and second
  law of quantum thermodynamics: A consistent derivation based on a microscopic
  definition of entropy},}\ }\href {\doibase 10.1103/PRXQuantum.2.030202}
  {\bibfield  {journal} {\bibinfo  {journal} {PRX Quantum}\ }\textbf {\bibinfo
  {volume} {2}},\ \bibinfo {pages} {030202} (\bibinfo {year}
  {2021}{\natexlab{a}})}\BibitemShut {NoStop}%
\bibitem [{\citenamefont {Jiang}(2014{\natexlab{a}})}]{JiangPRE}%
  \BibitemOpen
  \bibfield  {author} {\bibinfo {author} {\bibfnamefont {J.-H.}\ \bibnamefont
  {Jiang}},\ }\bibfield  {title} {\enquote {\bibinfo {title} {Thermodynamic
  bounds and general properties of optimal efficiency and power in linear
  responses},}\ }\href {\doibase 10.1103/PhysRevE.90.042126} {\bibfield
  {journal} {\bibinfo  {journal} {Phys. Rev. E}\ }\textbf {\bibinfo {volume}
  {90}},\ \bibinfo {pages} {042126} (\bibinfo {year}
  {2014}{\natexlab{a}})}\BibitemShut {NoStop}%
\bibitem [{\citenamefont {Snyder}\ and\ \citenamefont {Ursell}(2003)}]{Snyder}%
  \BibitemOpen
  \bibfield  {author} {\bibinfo {author} {\bibfnamefont {G.~J.}\ \bibnamefont
  {Snyder}}\ and\ \bibinfo {author} {\bibfnamefont {T.~S.}\ \bibnamefont
  {Ursell}},\ }\bibfield  {title} {\enquote {\bibinfo {title} {Thermoelectric
  efficiency and compatibility},}\ }\href {\doibase
  10.1103/PhysRevLett.91.148301} {\bibfield  {journal} {\bibinfo  {journal}
  {Phys. Rev. Lett.}\ }\textbf {\bibinfo {volume} {91}},\ \bibinfo {pages}
  {148301} (\bibinfo {year} {2003})}\BibitemShut {NoStop}%
\bibitem [{\citenamefont {Jiang}(2014{\natexlab{b}})}]{JiangJAP}%
  \BibitemOpen
  \bibfield  {author} {\bibinfo {author} {\bibfnamefont {J.-H.}\ \bibnamefont
  {Jiang}},\ }\bibfield  {title} {\enquote {\bibinfo {title} {Enhancing
  efficiency and power of quantum-dots resonant tunneling thermoelectrics in
  three-terminal geometry by cooperative effects},}\ }\href {\doibase
  10.1063/1.4901120} {\bibfield  {journal} {\bibinfo  {journal} {J. Appl.
  Phys.}\ }\textbf {\bibinfo {volume} {116}},\ \bibinfo {pages} {194303}
  (\bibinfo {year} {2014}{\natexlab{b}})}\BibitemShut {NoStop}%
\bibitem [{\citenamefont {Lu}\ \emph {et~al.}(2017)\citenamefont {Lu},
  \citenamefont {Wang}, \citenamefont {Liu},\ and\ \citenamefont
  {Jiang}}]{MyJAP}%
  \BibitemOpen
  \bibfield  {author} {\bibinfo {author} {\bibfnamefont {J.}~\bibnamefont
  {Lu}}, \bibinfo {author} {\bibfnamefont {R.}~\bibnamefont {Wang}}, \bibinfo
  {author} {\bibfnamefont {Y.}~\bibnamefont {Liu}}, \ and\ \bibinfo {author}
  {\bibfnamefont {J.-H.}\ \bibnamefont {Jiang}},\ }\bibfield  {title} {\enquote
  {\bibinfo {title} {Thermoelectric cooperative effect in three-terminal
  elastic transport through a quantum dot},}\ }\href {\doibase
  10.1063/1.4995532} {\bibfield  {journal} {\bibinfo  {journal} {J. Appl.
  Phys.}\ }\textbf {\bibinfo {volume} {122}},\ \bibinfo {pages} {044301}
  (\bibinfo {year} {2017})}\BibitemShut {NoStop}%
\bibitem [{\citenamefont {Liu}\ \emph {et~al.}(2020)\citenamefont {Liu},
  \citenamefont {Lu}, \citenamefont {Wang}, \citenamefont {Wang},\ and\
  \citenamefont {Jiang}}]{CPB}%
  \BibitemOpen
  \bibfield  {author} {\bibinfo {author} {\bibfnamefont {Y.}~\bibnamefont
  {Liu}}, \bibinfo {author} {\bibfnamefont {J.}~\bibnamefont {Lu}}, \bibinfo
  {author} {\bibfnamefont {R.}~\bibnamefont {Wang}}, \bibinfo {author}
  {\bibfnamefont {C.}~\bibnamefont {Wang}}, \ and\ \bibinfo {author}
  {\bibfnamefont {J.-H.}\ \bibnamefont {Jiang}},\ }\bibfield  {title} {\enquote
  {\bibinfo {title} {Energy cooperation in quantum thermoelectric systems
  withmultiple electric currents},}\ }\href {\doibase 10.1088/1674-1056/ab7da5}
  {\bibfield  {journal} {\bibinfo  {journal} {Chin. Phys. B}\ }\textbf
  {\bibinfo {volume} {29}},\ \bibinfo {eid} {40504} (\bibinfo {year}
  {2020})}\BibitemShut {NoStop}%
\bibitem [{\citenamefont {Jiang}\ \emph {et~al.}(2015)\citenamefont {Jiang},
  \citenamefont {Kulkarni}, \citenamefont {Segal},\ and\ \citenamefont
  {Imry}}]{Jiangtransistors}%
  \BibitemOpen
  \bibfield  {author} {\bibinfo {author} {\bibfnamefont {J.-H.}\ \bibnamefont
  {Jiang}}, \bibinfo {author} {\bibfnamefont {M.}~\bibnamefont {Kulkarni}},
  \bibinfo {author} {\bibfnamefont {D.}~\bibnamefont {Segal}}, \ and\ \bibinfo
  {author} {\bibfnamefont {Y.}~\bibnamefont {Imry}},\ }\bibfield  {title}
  {\enquote {\bibinfo {title} {Phonon thermoelectric transistors and
  rectifiers},}\ }\href {\doibase 10.1103/PhysRevB.92.045309} {\bibfield
  {journal} {\bibinfo  {journal} {Phys. Rev. B}\ }\textbf {\bibinfo {volume}
  {92}},\ \bibinfo {pages} {045309} (\bibinfo {year} {2015})}\BibitemShut
  {NoStop}%
\bibitem [{\citenamefont {Lu}\ \emph {et~al.}(2020)\citenamefont {Lu},
  \citenamefont {Wang}, \citenamefont {Wang},\ and\ \citenamefont
  {Jiang}}]{MyPRBtransistor}%
  \BibitemOpen
  \bibfield  {author} {\bibinfo {author} {\bibfnamefont {J.}~\bibnamefont
  {Lu}}, \bibinfo {author} {\bibfnamefont {R.}~\bibnamefont {Wang}}, \bibinfo
  {author} {\bibfnamefont {C.}~\bibnamefont {Wang}}, \ and\ \bibinfo {author}
  {\bibfnamefont {J.-H.}\ \bibnamefont {Jiang}},\ }\bibfield  {title} {\enquote
  {\bibinfo {title} {Brownian thermal transistors and refrigerators in
  mesoscopic systems},}\ }\href {\doibase 10.1103/PhysRevB.102.125405}
  {\bibfield  {journal} {\bibinfo  {journal} {Phys. Rev. B}\ }\textbf {\bibinfo
  {volume} {102}},\ \bibinfo {pages} {125405} (\bibinfo {year}
  {2020})}\BibitemShut {NoStop}%
\bibitem [{\citenamefont {Chen}(2005)}]{gchen2005book}%
  \BibitemOpen
  \bibfield  {author} {\bibinfo {author} {\bibfnamefont {G.}~\bibnamefont
  {Chen}},\ }\href@noop {} {\emph {\bibinfo {title} {Nanoscale Energy Transport
  and Conversion}}}\ (\bibinfo  {publisher} {Oxford University Press, London},\
  \bibinfo {year} {2005})\BibitemShut {NoStop}%
\bibitem [{\citenamefont {Datta}(1995)}]{datta}%
  \BibitemOpen
  \bibfield  {author} {\bibinfo {author} {\bibfnamefont {S.}~\bibnamefont
  {Datta}},\ }\href@noop {} {\emph {\bibinfo {title} {Electronic transport in
  mesoscopic systems}}}\ (\bibinfo  {publisher} {Cambridge university press,
  Cambridge, UK},\ \bibinfo {year} {1995})\BibitemShut {NoStop}%
\bibitem [{\citenamefont {Esposito}\ \emph {et~al.}(2009)\citenamefont
  {Esposito}, \citenamefont {Lindenberg},\ and\ \citenamefont {den
  Broeck}}]{Esposito09}%
  \BibitemOpen
  \bibfield  {author} {\bibinfo {author} {\bibfnamefont {M.}~\bibnamefont
  {Esposito}}, \bibinfo {author} {\bibfnamefont {K.}~\bibnamefont
  {Lindenberg}}, \ and\ \bibinfo {author} {\bibfnamefont {C.~Van}\ \bibnamefont
  {den Broeck}},\ }\bibfield  {title} {\enquote {\bibinfo {title}
  {Thermoelectric efficiency at maximum power in a quantum dot},}\ }\href
  {\doibase 10.1209/0295-5075/85/60010} {\bibfield  {journal} {\bibinfo
  {journal} {Europhys. Lett.}\ }\textbf {\bibinfo {volume} {85}},\ \bibinfo
  {pages} {60010} (\bibinfo {year} {2009})}\BibitemShut {NoStop}%
\bibitem [{\citenamefont {Kedem}\ and\ \citenamefont {Caplan}(1965)}]{Kedem}%
  \BibitemOpen
  \bibfield  {author} {\bibinfo {author} {\bibfnamefont {O.}~\bibnamefont
  {Kedem}}\ and\ \bibinfo {author} {\bibfnamefont {S.~R.}\ \bibnamefont
  {Caplan}},\ }\bibfield  {title} {\enquote {\bibinfo {title} {Degree of
  coupling and its relation to efficiency of energy conversion},}\ }\href
  {\doibase 10.1039/TF9656101897} {\bibfield  {journal} {\bibinfo  {journal}
  {Trans. Faraday Soc.}\ }\textbf {\bibinfo {volume} {61}},\ \bibinfo {pages}
  {1897--1911} (\bibinfo {year} {1965})}\BibitemShut {NoStop}%
\bibitem [{\citenamefont {Saryal}\ \emph {et~al.}(2021)\citenamefont {Saryal},
  \citenamefont {Gerry}, \citenamefont {Khait}, \citenamefont {Segal},\ and\
  \citenamefont {Agarwalla}}]{BijayPRL21}%
  \BibitemOpen
  \bibfield  {author} {\bibinfo {author} {\bibfnamefont {S.}~\bibnamefont
  {Saryal}}, \bibinfo {author} {\bibfnamefont {M.}~\bibnamefont {Gerry}},
  \bibinfo {author} {\bibfnamefont {I.}~\bibnamefont {Khait}}, \bibinfo
  {author} {\bibfnamefont {D.}~\bibnamefont {Segal}}, \ and\ \bibinfo {author}
  {\bibfnamefont {B.~K.}\ \bibnamefont {Agarwalla}},\ }\bibfield  {title}
  {\enquote {\bibinfo {title} {Universal bounds on fluctuations in continuous
  thermal machines},}\ }\href {\doibase 10.1103/PhysRevLett.127.190603}
  {\bibfield  {journal} {\bibinfo  {journal} {Phys. Rev. Lett.}\ }\textbf
  {\bibinfo {volume} {127}},\ \bibinfo {pages} {190603} (\bibinfo {year}
  {2021})}\BibitemShut {NoStop}%
\bibitem [{\citenamefont {Strasberg}\ and\ \citenamefont
  {Winter}(2021{\natexlab{b}})}]{PRXQuantum}%
  \BibitemOpen
  \bibfield  {author} {\bibinfo {author} {\bibfnamefont {P.}~\bibnamefont
  {Strasberg}}\ and\ \bibinfo {author} {\bibfnamefont {A.}~\bibnamefont
  {Winter}},\ }\bibfield  {title} {\enquote {\bibinfo {title} {First and second
  law of quantum thermodynamics: A consistent derivation based on a microscopic
  definition of entropy},}\ }\href {\doibase 10.1103/PRXQuantum.2.030202}
  {\bibfield  {journal} {\bibinfo  {journal} {PRX Quantum}\ }\textbf {\bibinfo
  {volume} {2}},\ \bibinfo {pages} {030202} (\bibinfo {year}
  {2021}{\natexlab{b}})}\BibitemShut {NoStop}%
\bibitem [{\citenamefont {Holubec}\ and\ \citenamefont
  {Ryabov}(2018)}]{HolubecPRL}%
  \BibitemOpen
  \bibfield  {author} {\bibinfo {author} {\bibfnamefont {V.}~\bibnamefont
  {Holubec}}\ and\ \bibinfo {author} {\bibfnamefont {A.}~\bibnamefont
  {Ryabov}},\ }\bibfield  {title} {\enquote {\bibinfo {title} {Cycling tames
  power fluctuations near optimum efficiency},}\ }\href {\doibase
  10.1103/PhysRevLett.121.120601} {\bibfield  {journal} {\bibinfo  {journal}
  {Phys. Rev. Lett.}\ }\textbf {\bibinfo {volume} {121}},\ \bibinfo {pages}
  {120601} (\bibinfo {year} {2018})}\BibitemShut {NoStop}%
\bibitem [{\citenamefont {Van~den Broeck}(2005)}]{Van2005}%
  \BibitemOpen
  \bibfield  {author} {\bibinfo {author} {\bibfnamefont {C.}~\bibnamefont
  {Van~den Broeck}},\ }\bibfield  {title} {\enquote {\bibinfo {title}
  {Thermodynamic efficiency at maximum power},}\ }\href {\doibase
  10.1103/PhysRevLett.95.190602} {\bibfield  {journal} {\bibinfo  {journal}
  {Phys. Rev. Lett.}\ }\textbf {\bibinfo {volume} {95}},\ \bibinfo {pages}
  {190602} (\bibinfo {year} {2005})}\BibitemShut {NoStop}%
\bibitem [{\citenamefont {Proesmans}\ \emph
  {et~al.}(2016{\natexlab{b}})\citenamefont {Proesmans}, \citenamefont
  {Cleuren},\ and\ \citenamefont {Van~den Broeck}}]{PED}%
  \BibitemOpen
  \bibfield  {author} {\bibinfo {author} {\bibfnamefont {K.}~\bibnamefont
  {Proesmans}}, \bibinfo {author} {\bibfnamefont {B.}~\bibnamefont {Cleuren}},
  \ and\ \bibinfo {author} {\bibfnamefont {C.}~\bibnamefont {Van~den Broeck}},\
  }\bibfield  {title} {\enquote {\bibinfo {title} {Power-efficiency-dissipation
  relations in linear thermodynamics},}\ }\href {\doibase
  10.1103/PhysRevLett.116.220601} {\bibfield  {journal} {\bibinfo  {journal}
  {Phys. Rev. Lett.}\ }\textbf {\bibinfo {volume} {116}},\ \bibinfo {pages}
  {220601} (\bibinfo {year} {2016}{\natexlab{b}})}\BibitemShut {NoStop}%
\bibitem [{\citenamefont {Brandner}\ \emph {et~al.}(2015)\citenamefont
  {Brandner}, \citenamefont {Saito},\ and\ \citenamefont
  {Seifert}}]{BrandnerPRX}%
  \BibitemOpen
  \bibfield  {author} {\bibinfo {author} {\bibfnamefont {K.}~\bibnamefont
  {Brandner}}, \bibinfo {author} {\bibfnamefont {K.}~\bibnamefont {Saito}}, \
  and\ \bibinfo {author} {\bibfnamefont {U.}~\bibnamefont {Seifert}},\
  }\bibfield  {title} {\enquote {\bibinfo {title} {Thermodynamics of micro- and
  nano-systems driven by periodic temperature variations},}\ }\href {\doibase
  10.1103/PhysRevX.5.031019} {\bibfield  {journal} {\bibinfo  {journal} {Phys.
  Rev. X}\ }\textbf {\bibinfo {volume} {5}},\ \bibinfo {pages} {031019}
  (\bibinfo {year} {2015})}\BibitemShut {NoStop}%
\bibitem [{\citenamefont {Lu}\ \emph {et~al.}(2019{\natexlab{b}})\citenamefont
  {Lu}, \citenamefont {Liu}, \citenamefont {Wang}, \citenamefont {Wang},\ and\
  \citenamefont {Jiang}}]{trade-off}%
  \BibitemOpen
  \bibfield  {author} {\bibinfo {author} {\bibfnamefont {J.}~\bibnamefont
  {Lu}}, \bibinfo {author} {\bibfnamefont {Y.}~\bibnamefont {Liu}}, \bibinfo
  {author} {\bibfnamefont {R.}~\bibnamefont {Wang}}, \bibinfo {author}
  {\bibfnamefont {C.}~\bibnamefont {Wang}}, \ and\ \bibinfo {author}
  {\bibfnamefont {J.-H.}\ \bibnamefont {Jiang}},\ }\bibfield  {title} {\enquote
  {\bibinfo {title} {Optimal efficiency and power, and their trade-off in
  three-terminal quantum thermoelectric engines with two output electric
  currents},}\ }\href {\doibase 10.1103/PhysRevB.100.115438} {\bibfield
  {journal} {\bibinfo  {journal} {Phys. Rev. B}\ }\textbf {\bibinfo {volume}
  {100}},\ \bibinfo {pages} {115438} (\bibinfo {year}
  {2019}{\natexlab{b}})}\BibitemShut {NoStop}%
\bibitem [{\citenamefont {Whitney}\ and\ \citenamefont
  {Saito}(2019)}]{WhitneySciPost}%
  \BibitemOpen
  \bibfield  {author} {\bibinfo {author} {\bibfnamefont {R.~S.}\ \bibnamefont
  {Whitney}}\ and\ \bibinfo {author} {\bibfnamefont {K.}~\bibnamefont
  {Saito}},\ }\bibfield  {title} {\enquote {\bibinfo {title} {{Thermoelectric
  coefficients and the figure of merit for large open quantum dots}},}\ }\href
  {\doibase 10.21468/SciPostPhys.6.1.012} {\bibfield  {journal} {\bibinfo
  {journal} {SciPost Phys.}\ }\textbf {\bibinfo {volume} {6}},\ \bibinfo
  {pages} {12} (\bibinfo {year} {2019})}\BibitemShut {NoStop}%
\bibitem [{\citenamefont {Niedenzu}\ and\ \citenamefont
  {Kurizki}(2018)}]{Niedenzu18NJP}%
  \BibitemOpen
  \bibfield  {author} {\bibinfo {author} {\bibfnamefont {W.}~\bibnamefont
  {Niedenzu}}\ and\ \bibinfo {author} {\bibfnamefont {G.}~\bibnamefont
  {Kurizki}},\ }\bibfield  {title} {\enquote {\bibinfo {title} {Cooperative
  many-body enhancement of quantum thermal machine power},}\ }\href {\doibase
  10.1088/1367-2630/aaed55} {\bibfield  {journal} {\bibinfo  {journal} {New J.
  Phys.}\ }\textbf {\bibinfo {volume} {20}},\ \bibinfo {pages} {113038}
  (\bibinfo {year} {2018})}\BibitemShut {NoStop}%
\bibitem [{\citenamefont {Lu}\ \emph {et~al.}(2019{\natexlab{c}})\citenamefont
  {Lu}, \citenamefont {Zhuo}, \citenamefont {Sun},\ and\ \citenamefont
  {Jiang}}]{CooperativeSpin}%
  \BibitemOpen
  \bibfield  {author} {\bibinfo {author} {\bibfnamefont {J.~C.}\ \bibnamefont
  {Lu}}, \bibinfo {author} {\bibfnamefont {F.~J.}\ \bibnamefont {Zhuo}},
  \bibinfo {author} {\bibfnamefont {Z.~Z.}\ \bibnamefont {Sun}}, \ and\
  \bibinfo {author} {\bibfnamefont {J.~H.}\ \bibnamefont {Jiang}},\ }\bibfield
  {title} {\enquote {\bibinfo {title} {Cooperative spin caloritronic
  devices},}\ }\href {\doibase 10.30919/esee8c353} {\bibfield  {journal}
  {\bibinfo  {journal} {ES Energy. Environ.}\ }\textbf {\bibinfo {volume}
  {7}},\ \bibinfo {pages} {17--28} (\bibinfo {year}
  {2019}{\natexlab{c}})}\BibitemShut {NoStop}%
\bibitem [{\citenamefont {Lu}\ \emph {et~al.}(2021)\citenamefont {Lu},
  \citenamefont {Jiang},\ and\ \citenamefont {Imry}}]{MyPRBdemon}%
  \BibitemOpen
  \bibfield  {author} {\bibinfo {author} {\bibfnamefont {J.}~\bibnamefont
  {Lu}}, \bibinfo {author} {\bibfnamefont {J.-H.}\ \bibnamefont {Jiang}}, \
  and\ \bibinfo {author} {\bibfnamefont {Y.}~\bibnamefont {Imry}},\ }\bibfield
  {title} {\enquote {\bibinfo {title} {Unconventional four-terminal
  thermoelectric transport due to inelastic transport: Cooling by transverse
  heat current, transverse thermoelectric effect, and maxwell demon},}\ }\href
  {\doibase 10.1103/PhysRevB.103.085429} {\bibfield  {journal} {\bibinfo
  {journal} {Phys. Rev. B}\ }\textbf {\bibinfo {volume} {103}},\ \bibinfo
  {pages} {085429} (\bibinfo {year} {2021})}\BibitemShut {NoStop}%
\bibitem [{\citenamefont {Whitney}\ \emph {et~al.}(2016)\citenamefont
  {Whitney}, \citenamefont {S\'anchez}, \citenamefont {Haupt},\ and\
  \citenamefont {Splettstoesser}}]{WhitneyPhysE}%
  \BibitemOpen
  \bibfield  {author} {\bibinfo {author} {\bibfnamefont {R.~S.}\ \bibnamefont
  {Whitney}}, \bibinfo {author} {\bibfnamefont {R.}~\bibnamefont {S\'anchez}},
  \bibinfo {author} {\bibfnamefont {F.}~\bibnamefont {Haupt}}, \ and\ \bibinfo
  {author} {\bibfnamefont {J.}~\bibnamefont {Splettstoesser}},\ }\bibfield
  {title} {\enquote {\bibinfo {title} {Thermoelectricity without absorbing
  energy from the heat sources},}\ }\href {\doibase
  https://doi.org/10.1016/j.physe.2015.09.025} {\bibfield  {journal} {\bibinfo
  {journal} {Physica E}\ }\textbf {\bibinfo {volume} {75}},\ \bibinfo {pages}
  {257 -- 265} (\bibinfo {year} {2016})}\BibitemShut {NoStop}%
\bibitem [{\citenamefont {Xi}\ \emph {et~al.}(2021)\citenamefont {Xi},
  \citenamefont {Wang}, \citenamefont {Lu},\ and\ \citenamefont
  {Jiang}}]{MyCPL21}%
  \BibitemOpen
  \bibfield  {author} {\bibinfo {author} {\bibfnamefont {M.}~\bibnamefont
  {Xi}}, \bibinfo {author} {\bibfnamefont {R.}~\bibnamefont {Wang}}, \bibinfo
  {author} {\bibfnamefont {J.}~\bibnamefont {Lu}}, \ and\ \bibinfo {author}
  {\bibfnamefont {J.-H.}\ \bibnamefont {Jiang}},\ }\bibfield  {title} {\enquote
  {\bibinfo {title} {Coulomb thermoelectric drag in four-terminal mesoscopic
  quantum transport},}\ }\href {\doibase 10.1088/0256-307X/38/8/088801}
  {\bibfield  {journal} {\bibinfo  {journal} {Chin. Phys. Lett.}\ }\textbf
  {\bibinfo {volume} {38}},\ \bibinfo {eid} {088801} (\bibinfo {year}
  {2021})}\BibitemShut {NoStop}%
\end{thebibliography}%

\end{document}